%% file: main_PRX_quantum_v2.tex
\documentclass[prx,aps,showpacs,twocolumn,%
    preprintnumbers,amsmath,amssymb,superscriptaddress,longbibliography,%
    nofootinbib
    ,a4paper
    ]{revtex4-2}

\usepackage[english]{babel}
\usepackage{subfiles}
\usepackage{bbold}
\usepackage{amssymb}
\usepackage{comment}
\usepackage{microtype}
\usepackage[most]{tcolorbox}
\usepackage{amsmath}
\usepackage{amsthm}
\usepackage{ulem} 

\usepackage{comment}
\usepackage{chngpage}

\usepackage{amssymb}

\usepackage[most]{tcolorbox}

\usepackage{amsmath}

\usepackage{amsthm}

\usepackage{comment}
\usepackage[colorlinks,citecolor=blue,urlcolor=blue,linkcolor=magenta]{hyperref}
\usepackage[all]{hypcap} 

\usepackage{tikzit}
\input{ZX.tikzstyles}
\input{ZX.tikzdefs}

\newif\ifappendix
\appendixfalse
\appendixtrue 

\def\changemargin#1#2{\list{}{\rightmargin#2\leftmargin#1}\item[]}
\let\endchangemargin=\endlist 

\theoremstyle{plain}

\theoremstyle{definition}

\def\Plus{\texttt{+}}
\def\Minus{\texttt{-}}

\newcommand{\noteAG}[1]{{\color{magenta} [AG: #1]}}
\newcommand{\nick}[1]{\textsf{\color{blue}[NC:#1]}}
\newcommand{\richard}[1]{ \textsf{\color{purple}[RE:#1]}}
\newcommand{\john}[1]{\textsf{\color{green}[JvdW:#1]}}

\begin{document}

\title{AKLT-states as ZX-diagrams: diagrammatic reasoning for quantum states}

\author{Richard D. P. East}
\affiliation{Univ.~Grenoble Alpes, LIG, 38401 Saint-Martin-d'H\`eres, France}
\affiliation{Univ.~Grenoble Alpes, CNRS, Grenoble INP, Institut N\'eel, 38000 Grenoble, France}
\author{John van de Wetering}
\affiliation{Radboud University Nijmegen, The Netherlands}
\author{Nicholas Chancellor}
\affiliation{Durham University physics department and Durham-Newcastle Joint Quantum Centre, South Road, Durham UK}
\author{Adolfo G. Grushin}
\affiliation{Univ. Grenoble Alpes, CNRS, Grenoble INP, Institut N\'eel, 38000 Grenoble, France}

\begin{abstract}
From Feynman diagrams to tensor networks, diagrammatic representations of computations in quantum mechanics have catalysed progress in physics. These diagrams represent the underlying mathematical operations and aid physical interpretation, but cannot generally be computed with directly. 
In this paper we introduce the ZXH-calculus, a graphical language based on the ZX-calculus, that we use to represent and reason about many-body states entirely graphically.
As a demonstration, we express the 1D AKLT state, a symmetry protected topological state, in the ZXH-calculus by developing a representation of spins higher than 1/2 within the calculus. 
{ 
By exploiting the simplifying power of the ZXH-calculus rules we show how this representation straightforwardly recovers the AKLT matrix-product state representation, the existence of topologically protected edge states, and the non-vanishing of a string order parameter. Extending beyond these known properties, our diagrammatic approach also allows us to analytically derive that the Berry phase of any finite-length 1D AKLT chain is $\pi$.
} 
In addition, we provide an alternative proof that the 2D AKLT state on a hexagonal lattice can be reduced to a graph state, demonstrating that it is a universal quantum computing resource. 
{ 
Lastly, we build 2D higher-order topological phases diagrammatically, which we use to illustrate a symmetry-breaking phase transition.
}
Our results show that the ZXH-calculus is a powerful language for representing and computing with physical states entirely graphically, paving the way to develop more efficient many-body algorithms
{ and giving a novel diagrammatic perspective on quantum phase transitions}.
\end{abstract}

\date{\today}

\maketitle

\section{Introduction}

\begin{figure*}
	\centering
	\hspace{-18.3cm} 
	\includegraphics[width=1.0\textwidth]{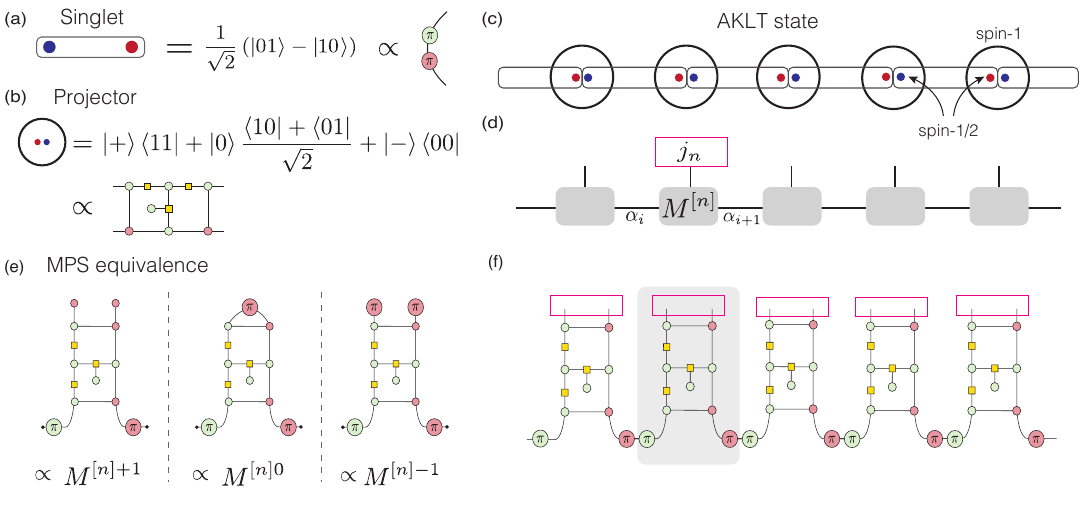}
	\caption{ZXH representation of the AKLT state. (a) and (b) show the singlet and symmetric projector and their ZXH representation. 
	These are the basic building blocks of the 1D AKLT state, shown pictorially in (c). 
	(d) gives the MPS representation of the 1D AKLT state, while
	(f) gives its ZXH representation, which consists of the components in (a) and (b).
	The shaded gray square in (f) highlights the part of the diagram from which one obtains the three MPS matrices $M^{[n]+1},M^{[n]0},$ and $M^{[n]-1}$ needed for the AKLT state. 
	The diagrams of these matrices are shown in (e), and are obtained by fixing the physical index (highlighted by the magenta rectangles in (f)).
	}
	\label{fig:AKLT-mps}
\end{figure*}

Representing involved mathematical formulae with simple diagrams has been a common strategy to drive progress in physics. An important and widespread example of this are Feynman diagrams~\cite{Kaiser05}, where the often cumbersome integrals that predict the amplitude of a quantum field theory process are ordered in perturbation theory with the aid of simple diagrammatic representations. 

A more recent example is the formulation of the quantum many-body problem in terms of tensor networks, that are often represented diagrammatically~\cite{Orus:2019vt,Cirac2020}. 
Tensor networks have triggered the development of computationally efficient variational algorithms that find an approximate solution to many-body problems~\cite{White1992,White1993,Verstraete2008,Schollwock11,10.21468/SciPostPhysLectNotes.5,Bridgeman_2017,Haegeman2017}. These formulations are based on efficient representations of quantum states, for which matrix product states~\cite{Fannes:1992dl,Ostlund1995,Rommer1997} (MPS) and projected entangled pair states~\cite{Verstraete04,Orus2014} (PEPS) are among the most successful approaches. These states are often represented diagrammatically as sites that connect to each other by legs that represent tensor contractions.

Despite their unquestionable success in addressing the quantum many-body problem, tensor networks have known limitations. For example, MPS are a one-dimensional (1D) representation of the wavefunction, which limit their scope, while PEPS cannot be contracted both efficiently and exactly~\cite{Schuch2007,Orus:2019vt}. 
Additionally, there are limitations to the type of states one can represent efficiently using existing tensor networks. MPS are well suited to describe gapped Hamiltonians in one and two dimensions, but are less suited for critical states and higher dimensions~\cite{Calabrese_2004,Pollmann2009,Tagliacozzo2008}. PEPS can handle both gapped and critical states and can be defined in any dimension~\cite{Verstraete06}, but representing certain states is challenging, notably states with chiral topological order~\cite{Wahl2013,Dubail2015,Poilblanc2015,Chen20}. Although many other tensor-network methods exist~\cite{Orus:2019vt,Cirac2020}, each tailored to solve different issues, finding novel ways to represent states is an ongoing challenge. 

In this work we present a diagrammatic representation of quantum states with which we can compute directly, in contrast to the typical graphical representation of a tensor network which is merely a representation of the underlying mathematical operations (the tensor contractions). 
We use \emph{ZX-diagrams}, a type of tensor network that comes equipped with a set of graphical rewrite rules known as the \emph{ZX-calculus}. 
The ZX-calculus was developed to better understand the foundations of quantum information and entanglement~\cite{coecke2008interacting,coecke2011interacting,CKbook}. It has seen use in quantum circuit optimisation~\cite{duncan2019graph,kissinger2020reducing,deBeaudrap2020Techniques,Cowtan2020phasegadget}, measurement-based quantum computation~\cite{duncan2010rewriting,kissinger2017MBQC,Backens2020extraction} and surface code lattice surgery~\cite{horsman2017surgery,magicFactories,deBeaudrap2020Paulifusion,hanks2019effective}.
The goal of this paper is to explore how the ZX-calculus can be used 
to represent quantum states, and to extract their useful physical properties.

The power of the ZX-calculus stems from the fact that we can simplify a given diagram without calculating its underlying matrix: the diagram is the calculation. The ZX-calculus is \emph{complete}, which means that any diagrams representing the same linear map can be transformed into one another entirely diagrammatically~\cite{SimonCompleteness,ng2017universal,HarnyAmarCompleteness,vilmarteulercompleteness,zh-calculus}.
ZX-diagrams are generated by a small set of generators that are symmetric tensors acting on a two-dimensional (i.e.~spin-$1/2$) Hilbert space.
While ZX-diagrams can in principle represent any linear map between qubits, some particularly canonical constructions are relatively hard to represent, in particular `AND-like' constructions that are especially relevant for this work. To remedy this problem, in 2018 the \emph{ZH-calculus} was introduced~\cite{zh-calculus}. It adds another generator to the ZX-calculus, and suitable rewrite rules to reason about it. In this paper we will develop and use a slight variation on the ZH-calculus that we dub the \emph{ZXH-calculus}. 

The question we address in this work is to what extent the ZXH-calculus can efficiently represent quantum states, and simplify operations on them. We find that the ZXH-calculus presents some advantages compared to existing formulations, and an evident potential for further advances. 
We demonstrate this by writing 1D and 2D AKLT states~\cite{AKLT87,Affleck:1988hl} as ZXH-diagrams. 
For the 1D AKLT state we show that the ZXH representation allows us to detect its string order graphically~\cite{Nijs89}.
We also map the ZXH representation of the 1D AKLT state to its MPS representation~\cite{10.21468/SciPostPhysLectNotes.5}, establishing a bridge between graphical calculi and MPS representations.
{ By exploiting the benefits of the diagrammatic calculus we derive that the Berry phase of the 1D AKLT state is $\pi$ for any finite chain-length~\cite{Hatsugai:2006hb}.}
To exemplify the power of the ZXH representation further, we prove entirely diagrammatically that the 2D AKLT state reduces to a graph state under a suitable set of measurements. This result, originally proved in Ref.~\cite{wei2011affleck} and independently in Ref.~\cite{miyake2011quantum}, can be used to show that the 2D AKLT state is a universal resource for quantum computation. 
While Ref.~\cite{wei2011affleck} proved the reduction to a graph state using reasoning specific to their construction, using our representation it follows directly using relatively simple and standard diagrammatic rewrites of the ZXH-calculus. 

{ Lastly, we consider how crystalline symmetries can be implemented in ZXH by constructing a higher-order topological phase protected by mirror symmetry. We find that symmetric diagrams represent symmetric states, offering a straightforward, diagrammatic way to implement crystal symmetries, not available to other tensor networks. We show how this result is advantageous to describe phase transitions diagrammatically; by breaking the mirror-symmetries that protect the higher-order topological state down to four-fold rotations ($C_4$), we can observe how the topological end-modes, originally pinned to the corners, move along the boundary.}

For several of our computations we have used the \textsc{Python} software package \textsc{PyZX} to assist in the diagrammatic reasoning~\cite{kissinger2019Pyzx}. Many of the computations we present in this paper are shown for pedagogical purposes only as they can be performed in an entirely automated manner by \textsc{PyZX}, evidencing the power of using the ZXH-calculus to represent these states. For these calculations see the accompanying Jupyter notebooks\footnote{You can find the accompanying Jupyter notebooks \href{https://github.com/Quantomatic/pyzx/tree/4837ea92ec56a98af268401a2c3fcb32946d5faa/demos/AKLT}{here}.}.

Based on the early work on the ZX-calculus of~\cite{MichaelMasters,coecke2010compositional}, the authors of~\cite{biamonte2011categorical,denny2011algebraically} also graphically calculated properties of some tensor network states. However, they restricted to representing networks that are stabiliser states, and hence for which it is already known they can be efficiently contracted~\cite{aaronsongottesman2004}. In contrast, our work deals with states that are computationally universal~\cite{Brennen2008,wei2011affleck,miyake2011quantum}. 
Other related work is the quon graphical language~\cite{liu2017quon,liu2019quantized,jaffe2017constructive,jaffe2018holographic}, that has so far also focused on stabiliser protocols, and Ref.~\cite{Bauer2020} which recently developed a graphical tensor-network representation of path integrals describing topological phases.

The main difficulty in using the ZXH-calculus to represent arbitrary quantum states is that all the indices of the tensors in a ZX-diagram are of dimension two (i.e.~they are spin-1/2 degrees of freedom). 
Hence, to use ZXH-diagrams to represent quantum states that live in larger Hilbert spaces (such as the spin-1 states in a 1D AKLT state) we need to encode these larger Hilbert spaces into tensor products of two-dimensional Hilbert spaces. 
We solve this problem by resorting to the representation theory of $SU(2)$, which tells us there is a unique $N$-dimensional representation given by the symmetric subspace of $N-1$ copies of $\C^2$. 
Our construction of this symmetriser in terms of simple tensors and its ties to the representation theory of $SU(2)$ might be of broader interest.

As the intersection of readers familiar with both the ZX-calculus and the AKLT state is probably quite narrow, we give a self-contained introduction to both. In Section~\ref{sec:AKLTintro} we describe briefly what AKLT states are by introducing the paradigmatic 1D AKLT state. 
In Section~\ref{sec:intro to ZXH} we present a concise review of the ZX-calculus and its extension to the ZXH-calculus. 
Then in Section~\ref{sec:AKLT-in-ZXH} we represent the 1D AKLT state in the ZXH-calculus and demonstrate some calculations on it.
In Section~\ref{sec:higher-spins} we discuss how we can represent higher spin systems in the ZXH-calculus, and we use this in Section~\ref{sec:2DAKLT} to represent the 2D AKLT state on a hexagonal lattice in the ZXH-calculus and to derive its reduction to a graph state fully diagrammatically. 
We study symmetry transitions of states in Section~\ref{section:symmeries} and
we end with some concluding remarks in Section~\ref{sec:conclusion}.

\section{Preliminaries}

\subsection{Introduction to AKLT states \label{sec:AKLTintro}}

The one-dimensional AKLT Hamiltonian, named after Affleck, Lieb, Kennedy and Tasaki, is defined as~\cite{AKLT87}
\begin{equation}
\label{eq:AKLTham}
    H = \sum_{i} \vec{S}_i \vec{S}_{i+1} + \beta  (\vec{S}_i \vec{S}_{i+1})^2,
\end{equation}
where $\beta=1/3$. This Hamiltonian acts on a chain of $N$ spin-1 degrees of freedom. 
Hence, the local Hilbert space at each site is $\mathbb{C}^{3}$, on which we act with the spin operator $\vec{S}_i=(S^{x}_i,S^{y}_i,S^{z}_i)$, where the $S^{a}_i$ are the $3\times 3$ spin-1 matrices (these matrices, along with other additional information on the AKLT state is given in Appendix \ref{app:spin}). 
Using representation theory it can be shown that the Hilbert space of a chain with $N$ sites, $(\mathbb{C}^{3})^{\otimes N}$, can be represented by $N$ copies of the symmetric subspace of a pair of spin-1/2 particles. This decomposition is convenient for finding the groundstate of the AKLT Hamiltonian Eq.~\eqref{eq:AKLTham}, because this Hamiltonian can be written as a positive sum of spin $s=2$ projectors on neighbouring sites. 
Hence, by finding a state where two neighbouring spins are not in the $s=2$ subspace, we can construct the ground state of the AKLT Hamiltonian. 
Specifically, the groundstate can be constructed by decomposing each spin-1 site into two spin-1/2 sites that form singlets between neighbouring sites (Fig.~\ref{fig:AKLT-mps} (a)), and thus have a maximum $s=1$. These two spin-1/2 sites are then projected back to the physical $s=1$ at each site by the appropriate symmetrising projectors (Fig.~\ref{fig:AKLT-mps}(b)). 
By construction, the resulting state, depicted in Fig.~\ref{fig:AKLT-mps}(c) is annihilated by the $s=2$ projectors, and is therefore an exact ground state of Eq.~\eqref{eq:AKLTham}. 
We refer to this groundstate as the \emph{AKLT state}\footnote{With periodic boundary conditions the groundstate of the AKLT Hamiltonian is unique, but with open boundary conditions it is four-fold degenerate. When referring to `the AKLT state' we don't distinguish open or boundary conditions, but rather mean all these possible states, as is common practice in the literature.}.

{The AKLT state has three important properties that we will express using the ZXH-calculus~\cite{Nijs89,Hatsugai:2006hb,Gu2009,Pollmann2010}}. The first property stems from the fact that terminating the chain necessarily breaks two singlets, one at each edge, leaving two free spin-1/2 degrees of freedom at the edges. 
Since each spin-1/2 has a local Hilbert space of $\mathbb{C}^2$ (the dimensions corresponding to spin up or spin down), the AKLT state with open boundary conditions has a degeneracy of four ($2^2$).

The second property that we wish to express using the ZXH-calculus is that the AKLT state has a \emph{string order}~\cite{Nijs89}. 
Namely, the AKLT state is a superposition of all spin configurations where, if we ignore the spins with $s_z=0$, the remaining spins are ordered anti-ferromagnetically: a spin $s_z={\pm1}$ is followed by $s_z=\mp1$~\cite{Kim00}. 
For example, $\left|j_1,j_2,\cdots j_{N}\right\rangle = \left|1,0,0,0,-1,0,0,0,1\right\rangle$ is an allowed configuration, while $\left|1,0,0,1,0,0,0,1\right\rangle$ is not. 
Analogous to how a spin-1/2 antiferromagnetic order can be captured by an alternating spin-spin correlation function, this string order can be captured by defining a \emph{string order parameter}~\cite{Nijs89}.

{ The last property of the 1D AKLT chain that we consider in this work is its Berry phase.  In describing symmetry-protected topological phases, such as the 1D AKLT state, and their phase transitions, it is useful to define quantities that are quantised due to the underlying symmetries that protect the phase. Hatsugai showed that the Berry phase, to be defined in Sec.~\ref{sec:Berry-phase}, is quantised to $\pi$ for the 1D AKLT state in the thermodynamic limit~\cite{Hatsugai:2006hb}. 
This distinguishes this state from trivial 1D states, for which the Berry phase is zero. The quantisation of the Berry phase was later generalised to describe other symmetry protected topological phases~\cite{Hatsugai_2007,Hirano2008,Motoyama2008,Kariyado:2018bc}.} 

The AKLT state has also a simple MPS representation. To describe it we follow the notation of Ref.~\cite{10.21468/SciPostPhysLectNotes.5} from which we recall that any quantum state can be written as a product of matrices as
\begin{equation}
\label{eq:MPSWF}
    \scalebox{0.93}{$
     \left|\psi\right\rangle = \sum_{j_1,\cdots j_N}  \sum_{\alpha_2,\cdots \alpha_{N}} M^{[1]j_1}_{\alpha_1,\alpha_2}\textbf{}\cdots M^{[N]j_N}_{\alpha_N,\alpha_{N+1}}  \left|j_1,j_2,\cdots j_{N}\right\rangle.$} 
\end{equation}
The indices $j_{i}$ are called \emph{physical indices} because they span the local Hilbert space at a given site $n$ (e.g. $j_{i}=0,\pm1$ for spin-1). For a given $j_{i}$ and $n$ the $M^{[n]j_n}_{\alpha_{i}\alpha_{i+1}}$ are 
matrices in the indices $\alpha_{i}$, known as bond indices\footnote{Note that we do not sum over $\alpha_1$ and $\alpha_{N+1}$. For periodic boundary conditions one sets $\alpha_1=\alpha_{N+1}$ and sums over them.}. 
Although \eqref{eq:MPSWF} is an exact representation of any state of a finite system, the maximum dimension of the bond indices needed to write a given state, known as the bond dimension
 $\chi$, generally grows exponentially with system size. The bond dimension $\chi$ is a measure of the entanglement of the state we wish to represent~\cite{10.21468/SciPostPhysLectNotes.5}. 

The AKLT state can be written as an exact MPS of bond dimension $\chi=2$. The local Hilbert space of each site consists of three spin-$1$ states and, with periodic boundary conditions, each site is equivalent. The AKLT state is defined by the three matrices
\begin{eqnarray}
\nonumber
    M^{[n]+1}&=&\sqrt{\dfrac{2}{3}}
    \left(\begin{array}{cc}
      0   & 0 \\
      1   & 0
    \end{array}\right), \hspace{5mm} M^{[n]0}=\dfrac{1}{\sqrt{3}}
    \left(\begin{array}{cc}
      1  & 0 \\
      0   & -1
    \end{array}\right), \hspace{5mm}\\
    \label{eq:AKLTMPS}
    M^{[n]-1}&=&\sqrt{\dfrac{2}{3}}
    \left(\begin{array}{cc}
      0   & -1 \\
      0   & 0
    \end{array}\right).
\end{eqnarray}
which are the same for all sites $1<n<N$ in the bulk (see Fig.~\ref{fig:AKLT-mps}(d)).

The ideas behind the AKLT state and their generalisations are widely used to understand more complicated condensed matter systems~\cite{Anderson73}, and used as well as computational tools~\cite{Orus:2019vt}. The one-dimensional AKLT state can also be generalized to two-dimensional lattices~\cite{Affleck:1988hl}. The particular case we will consider in Section \ref{sec:2DAKLT} is the AKLT state on a hexagonal lattice with a spin-3/2 degree of freedom at each site. It can be constructed using entangled pairs of spin-1/2 states projected to the appropriate subspace. Hence the 2D AKLT state can be represented as a 2D PEPS with dimension $D=2$~\cite{Orus2014}. This state was shown to be a universal source for measurement-based quantum computation~\cite{wei2011affleck}.

\subsection{Introduction to the ZXH-calculus}
\label{sec:intro to ZXH}

\begin{figure*}
\centering
\hspace{-15.5cm}
\includegraphics[width=0.8\textwidth]{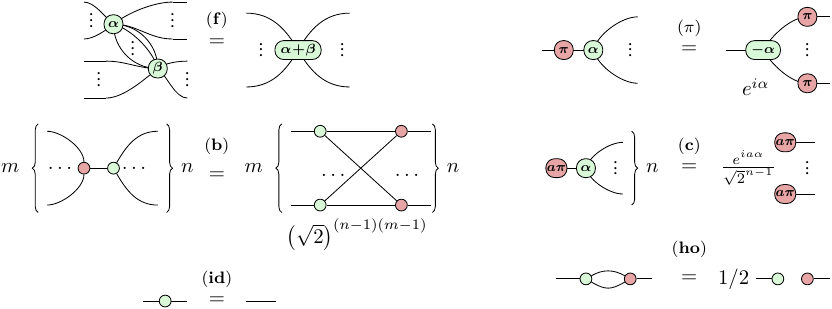}
\caption{
The rules of the ZX-calculus. These rules hold for all $\alpha, \beta \in [0, 2 \pi)$, and $a\in\{0,1\}$. They also hold with the colours red and green interchanged, and with inputs and outputs permuted freely. Note `...' should be read as `0 or more', hence the spiders on the left-hand side of \SpiderRule are connected by one or more wires. Furthermore, the addition in \SpiderRule{} is taken to be modulo $2\pi$. The right-hand side of \BialgRule is a fully-connected bipartite graph. The rulenames stand respectively for \SpiderRule{}use, \PiCopyRule{}opy, \BialgRule{}ialgebra, \CopyRule{}opy, \IdentityRule{}entity and \HopfRule{}pf.}
\label{fig:zx-rules}
\end{figure*}

In this paper we will use a graphical calculus that is a mixture of the ZX-calculus and the ZH-calculus. For ease of reference we dub this language the \emph{ZXH-calculus}.
First, we provide a brief overview of the more well-known ZX-calculus. For an in-depth
reference see Ref.~\cite{CKbook} or Ref.~\cite{vandewetering2020zxcalculus}.

The ZX-calculus is a diagrammatic language similar to 
quantum circuit notation~\cite{coecke2008interacting,coecke2011interacting}.  A \emph{ZX-diagram} (or simply
\emph{diagram}) consists of \emph{wires} and \emph{spiders}.  Wires
entering the diagram from the left are \emph{inputs}; wires exiting to
the right are \emph{outputs}.  Given two diagrams we can compose them
by joining the outputs of the first to the inputs of the second, or
form their tensor product by stacking the two diagrams.

Spiders are linear operations which can have any number of input or output
wires.  There are two varieties: Z-spiders depicted as green dots and X-spiders depicted as red dots, each of which can be labelled by a phase $\alpha\in\R$:
\begin{align}
\small
\begin{tikzpicture}
	\begin{pgfonlayer}{nodelayer}
		\node [style=Z phase dot] (0) at (0, 0) {$\alpha$};
		\node [style=none] (1) at (1.25, 1) {};
		\node [style=none] (2) at (-1.25, 1) {};
		\node [style=none] (3) at (-1.25, -1) {};
		\node [style=none] (4) at (1.25, -1) {};
		\node [style=none] (5) at (1.25, 0.5) {};
		\node [style=none] (6) at (-1.25, 0.5) {};
		\node [style=none, rotate=90] (7) at (-1, -0.25) {...};
		\node [style=none, rotate=90] (8) at (1, -0.25) {...};
	\end{pgfonlayer}
	\begin{pgfonlayer}{edgelayer}
		\draw [in=-141, out=0, looseness=0.75] (3.center) to (0);
		\draw [in=180, out=-39, looseness=0.75] (0) to (4.center);
		\draw [in=180, out=22, looseness=0.75] (0) to (5.center);
		\draw [in=180, out=39, looseness=0.75] (0) to (1.center);
		\draw [in=0, out=158, looseness=0.75] (0) to (6.center);
		\draw [in=141, out=0, looseness=0.75] (2.center) to (0);
	\end{pgfonlayer}
\end{tikzpicture}%
 \ &:= \ \ketbra{0\cdots 0}{0\cdots 0} +
e^{i \alpha} \ketbra{1\cdots 1}{1\cdots 1} \\
\begin{tikzpicture}
	\begin{pgfonlayer}{nodelayer}
		\node [style=X phase dot] (0) at (0, 0) {$\alpha$};
		\node [style=none] (1) at (1.25, 1) {};
		\node [style=none] (2) at (-1.25, 1) {};
		\node [style=none] (3) at (-1.25, -1) {};
		\node [style=none] (4) at (1.25, -1) {};
		\node [style=none] (5) at (1.25, 0.5) {};
		\node [style=none] (6) at (-1.25, 0.5) {};
		\node [style=none, rotate=90] (7) at (-1, -0.25) {...};
		\node [style=none, rotate=90] (8) at (1, -0.25) {...};
	\end{pgfonlayer}
	\begin{pgfonlayer}{edgelayer}
		\draw [in=-141, out=0, looseness=0.75] (3.center) to (0);
		\draw [in=180, out=-39, looseness=0.75] (0) to (4.center);
		\draw [in=180, out=22, looseness=0.75] (0) to (5.center);
		\draw [in=180, out=39, looseness=0.75] (0) to (1.center);
		\draw [in=0, out=158, looseness=0.75] (0) to (6.center);
		\draw [in=141, out=0, looseness=0.75] (2.center) to (0);
	\end{pgfonlayer}
\end{tikzpicture}%
 \ &:= \ \ketbra{+\cdots +}{+\cdots +} +
e^{i \alpha} \ketbra{-\cdots -}{-\cdots -}
\end{align}
Note that if you are reading this document in monochrome or otherwise have difficulty distinguishing green and red, Z-spiders will appear lightly-shaded and X-spiders darkly-shaded.
ZX-diagrams are constructed iteratively from these spiders by composing them either sequentially, which on the level of the linear map corresponds to the regular composition of linear maps, or by stacking them, which forms the tensor product of the linear maps they represent.
As a special case, diagrams with no inputs represent
(unnormalised) state preparations, while diagrams with no open wires represent complex scalars.

As a demonstration, let us write down some simple state preparations and unitaries in the ZX-calculus:
  \begin{align}
%
\ \  & = & \ketbra{+}{+} + e^{i \alpha} \ketbra{-}{-}\ \ & = & X_\alpha \label{eq:r-alph}
  \end{align}
  Note that, while \eqref{eq:g-alph} and \eqref{eq:r-alph} have a label $\alpha$, we have not given a label to the state preparations \eqref{eq:ket-+} and \eqref{eq:ket-0}. By convention, a spider without a label is taken to have a label of $0$. 
  When we take $\alpha=\pi$ in \eqref{eq:g-alph} and \eqref{eq:r-alph} we get Pauli matrices:
  \begin{equation}
%
%
\ \ =\ \  X \qquad\qquad
  \end{equation}

By composing spiders we can make more complicated linear maps, such as the CNOT gate:
\begin{equation}\label{diagram:cnot}
	\scalebox{1.0}{%
%
	\ \ \propto \ \ \text{CNOT}
\end{equation}
Here the symbol `$\propto$' denotes that the diagram is proportional to the gate, i.e.~that there exists a global non-zero scalar correction (in this case, the diagram needs to be multiplied by $\sqrt{2}$) that makes them exactly equal.
For many of the calculations in this paper, the exact scalar value will not be important. For clarity, we will in those cases drop scalars implicitly. As above, we will write $\propto$ in a diagrammatic derivation to denote that the diagrams are merely equal up to a non-zero scalar.

We can treat a ZX-diagram as a graphical depiction of a tensor network,
similar in style to the work of e.g.~Penrose~\cite{Penrose}. In this interpretation, a wire between two spiders denotes a tensor contraction.
As tensors, Z and X spiders can be written as follows:
\begin{align}
\left( \  %
\begin{tikzpicture}
	\begin{pgfonlayer}{nodelayer}
		\node [style=Z phase dot] (0) at (0, 0) {$\alpha$};
	\end{pgfonlayer}
\end{tikzpicture}%
 \  \right)_{i_1...i_m}^{j_1...j_n} & =
{\small \begin{cases}
1 & \textrm{ if } i_1 = ... = i_m = j_1 = ... = j_n = 0 \\  
e^{i \alpha} & \textrm{ if } i_1 = ... = i_m = j_1 = ... = j_n = 1 \\
0 & \textrm{ otherwise} 
\end{cases}}
\end{align}
\begin{equation}
\scalebox{1.0}{
    $
    \left( \  %
\begin{tikzpicture}
	\begin{pgfonlayer}{nodelayer}
		\node [style=X phase dot] (0) at (0, 0) {$\alpha$};
	\end{pgfonlayer}
\end{tikzpicture}%
 \  \right)_{i_1...i_m}^{j_1...j_n} =
    {\small \left(\frac{1}{\sqrt{2}}\right)^{n+m} \cdot 
    \begin{cases}
    1 + e^{i \alpha}\!\! &\!\! \textrm{ if } \bigoplus_\alpha i_\alpha \oplus \bigoplus_\beta j_\beta = 0 \\  
    1 - e^{i \alpha}\!\! &\!\! \textrm{ if } \bigoplus_\alpha i_\alpha \oplus \bigoplus_\beta j_\beta = 1
    \end{cases}}
    $
}
\end{equation}
where $i_\alpha, j_\beta$ range over $\{0,1\}$ and $\oplus$ is addition modulo~2.

ZX-diagrams have a number of symmetries that make them easy to work with. In particular, we can treat a ZX-diagram as an undirected (multi-)graph, so that we can move the vertices around in the plane, bending,
unbending, crossing, and uncrossing wires, as long as the connectivity
and the order of the inputs and outputs is maintained. These deformations of the diagram do not affect the linear map it represents.
Indeed, the reader might have noticed that in the CNOT diagram~\eqref{diagram:cnot} we drew a {vertical} wire without explaining whether this denotes an input or an output from the Z- and X-spider. We are warranted in drawing it this way because:
\begin{equation}
\begin{tikzpicture}
	\begin{pgfonlayer}{nodelayer}
		\node [style=Green Node] (0) at (-2.75, 0.75) {};
		\node [style=Red Node] (1) at (-2, -0.75) {};
		\node [style=none] (2) at (-3.5, 0.75) {};
		\node [style=none] (3) at (-1.25, 0.75) {};
		\node [style=none] (4) at (-3.5, -0.75) {};
		\node [style=none] (5) at (-1.25, -0.75) {};
		\node [style=Green Node] (6) at (2.75, 0.75) {};
		\node [style=Red Node] (7) at (2, -0.75) {};
		\node [style=none] (8) at (3.5, 0.75) {};
		\node [style=none] (9) at (1.25, 0.75) {};
		\node [style=none] (10) at (3.5, -0.75) {};
		\node [style=none] (11) at (1.25, -0.75) {};
		\node [style=none] (12) at (0, 0) {=};
	\end{pgfonlayer}
	\begin{pgfonlayer}{edgelayer}
		\draw (2.center) to (0);
		\draw (0) to (3.center);
		\draw (5.center) to (1);
		\draw (1) to (4.center);
		\draw (0) to (1);
		\draw (8.center) to (6);
		\draw (6) to (9.center);
		\draw (11.center) to (7);
		\draw (7) to (10.center);
		\draw (6) to (7);
	\end{pgfonlayer}
\end{tikzpicture}%
\end{equation}
Besides these topological symmetries, ZX-diagrams have a set of rewrite rules associated to them, collectively referred to as the \emph{ZX-calculus}.
See Figure~\ref{fig:zx-rules} for a set of these rules.
Note that these rules also hold with the Z- and X-spider interchanged (i.e.~with the colours flipped).
When doing diagrammatic derivations, we will often label the equalities with one of the rule names of Figure~\ref{fig:zx-rules}, such as \SpiderRule, to denote that rule was used there.

\begin{figure*}
\centering
\hspace{-17.0cm}
\includegraphics[width=0.9\textwidth]{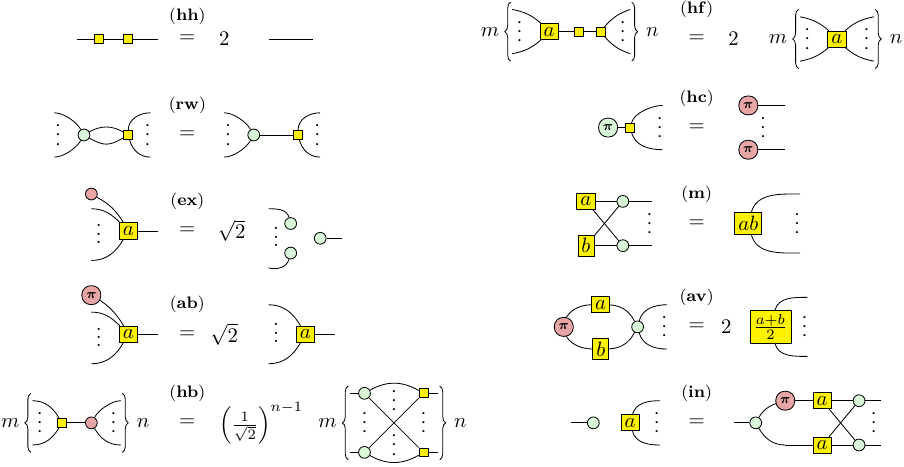}
\caption{
The rules of the ZH-calculus. These rules hold for all $a,b\in \C$. Note `...' should be read as `0 or more'. The right-hand side of \IntroRule and \HCompRule and the left-hand side of \MultRule contain fully connected bipartite graphs. 
In this paper we will only need the rules in the left column. The rest are shown for completeness.
The rule names stand for \HHRule{}-cancellation, \textbf{r}emove \textbf{w}ire, \ExplodeRule{}plode, \AbsorbRule{}sorb, \HCompRule{}ialgebra, \HSpiderRule{}use, \HCopyRule{}opy, \MultRule{}ultiplication, \AvgRule{}erage and \IntroRule{}troduction (as it introduces additional wires to the H-box on the left-hand side).
}
\label{fig:zh-rules}
\end{figure*}

In Figure~\ref{fig:zx-rules} we use a hybrid notation of writing numbers in the diagram itself to denote the correct global scalar needed to make the linear map of the two sides of the diagram exactly equal to one another. As noted above, we will sometimes drop these scalar factors when they are not relevant to the derivation at hand.

As a small demonstration of these rewrite rules, let us prove diagrammatically that the CNOT diagram~\eqref{diagram:cnot} indeed acts like the CNOT. The computational basis states are given by the following diagrams.
 \begin{equation}\label{diagram:basis-demo}
	\scalebox{1.0}{%
%
}
\end{equation}
Then we can check that the diagram has the correct action on these basis states:
\begin{equation}\label{diagram:cnot-demo}
	\scalebox{1.0}{%
%
%
}
\end{equation}

ZX-diagrams were introduced over a decade ago~\cite{coecke2008interacting} and have proven useful for reasoning about Clifford computation and single-qubit phase rotation gates~\cite{Backens1,duncan2019graph,horsman2017surgery}. 
It is however harder to reason about certain logical constructions, in particular the AND operation $\ket{x}\otimes \ket{y} \mapsto \ket{x\cdot y}$. {For instance, the only way to represent a CCNOT gate (also commonly known as the Toffoli gate) in the ZX-calculus is to expand it into Clifford and phase gates - which contains on the order of $\approx25$ spiders.}
In 2018 a new graphical calculus was introduced to remedy this problem: the \emph{ZH-calculus}~\cite{zh-calculus}. This calculus adds another generator to the ZX-calculus that allows for a compact representation of an AND gate.
This new generator is the \emph{H-box}:
\begin{equation}
\scalebox{0.9}{%
%
}\  := \ \sum a^{i_1\ldots i_m j_1\ldots j_n} \ket{j_1\ldots j_n}\bra{i_1\ldots i_m}
\end{equation}
Here $a$ can be any complex number, and the sum in this equation is over all $i_1,\ldots, i_m, j_1,\ldots, j_n\in\{0,1\}$ so that an H-box represents a matrix where all entries are equal to 1, except for the bottom right element, which is $a$. As a tensor we can write it as:
\begin{equation}
\left( \  %
%
%
 \  \right)_{i_1...i_m}^{j_1...j_n}\  =\ 
{\small \begin{cases}
a & \textrm{ if } i_1 = ... = i_m = j_1 = ... = j_n = 1 \\  
1 & \textrm{ otherwise} 
\end{cases}}
\end{equation}

Whereas for spiders we only draw the phase on the spider when it is nonzero, for H-boxes we only draw the label when it is not equal to $-1$. This is because the 1-input, 1-output H-box with a phase of $-1$ corresponds to the familiar Hadamard gate (up to a global scalar):
\begin{equation}\label{eq:Hdef}
%
\end{equation}
Note that in this paper we only need H-boxes labelled by~$-1$. We give the general definition for completeness' sake.

We have the following relations among the three generators, Z-spiders, X-spiders and H-boxes:
\begingroup
\allowdisplaybreaks
\begin{align}
	&\scalebox{0.95}{%
%
%
 \label{eq:H-to-ZX}
\end{align}
\endgroup
Note that it is also possible to represent H-boxes of higher \emph{arity}, i.e.~boxes with a larger number of input and output wires, using just Z- and X-spiders, but this is quite involved and not necessary for our purposes~\cite{ZHFourier}.

In addition to the rules of the ZX-calculus of Figure~\ref{fig:zx-rules} and the relations among the generators \eqref{eq:Z-to-X}--\eqref{eq:H-to-ZX} we also have some rules specific to the ZH-calculus; see Figure~\ref{fig:zh-rules}.
We present in Appendix~\ref{app:overview-rules} a condensed overview of all the rewrite rules and relations we have introduced so far.

An H-box with zero input and output wires that is labelled by $a$ is equal to the scalar $a$. This means we can always translate the scalars in the hybrid notation of Figures~\ref{fig:zx-rules} and~\ref{fig:zh-rules} into a ZH-diagram. For instance, the self-inverseness of the Hadamard gate can be represented as follows:
\begin{equation}
\begin{tikzpicture}
	\begin{pgfonlayer}{nodelayer}
		\node [style=none] (0) at (-5.5, 0) {};
		\node [style=none] (1) at (-1.75, 0) {};
		\node [style=H] (2) at (-4.5, 0) {};
		\node [style=H] (3) at (-3, 0) {};
		\node [style=none] (4) at (0, 0) {=};
		\node [style=none] (5) at (2.75, 0) {};
		\node [style=none] (6) at (4.75, 0) {};
		\node [style=H] (7) at (1.5, 0) {$2$};
	\end{pgfonlayer}
	\begin{pgfonlayer}{edgelayer}
		\draw (0.center) to (2);
		\draw (2) to (1.center);
		\draw (5.center) to (6.center);
	\end{pgfonlayer}
\end{tikzpicture}%
\end{equation}

ZH-diagrams are \emph{universal}, meaning that any linear map between complex vector spaces of dimension $2^n$ can be represented as a ZH-diagram.
Furthermore, the ZH-calculus is \emph{complete}, meaning that if two diagrams represent the same linear map, then we can find a sequence of rewrites from Figures~\ref{fig:zx-rules} and~\ref{fig:zh-rules} and equations~\eqref{eq:Z-to-X}--\eqref{eq:H-to-ZX} that transforms one diagram into the other~\cite{zh-calculus}.
However, in general, such a sequence of rewrites will involve diagrams of size exponential in the number of inputs and outputs (as otherwise we could establish efficient classical simulation of quantum computation, among other unlikely consequences such as P=NP).
The key to working with ZH-diagrams efficiently is then to find good heuristics for simplifying diagrams.

H-boxes allow us to straightforwardly represent controlled-phase gates. For instance, a CCZ($\theta$) gate, i.e.~a gate that maps the computational basis state $\ket{xyz}$ to $e^{i\theta xyz}\ket{xyz}$ is given by: 
\begin{equation}\label{eq:CCZ}

\end{equation}
As a special case of~\eqref{eq:CCZ} we also have the standard controlled-Z (CZ) gate:
\begin{equation}\label{eq:CZ}
%
\end{equation}
As another variation on these diagrams, we have the following diagram that we will use different iterations on throughout this paper:{
\begin{equation}\label{eq:zero-projector}
%
\end{equation}}
I.e.~this linear map throws away a $\ket{11}$ input, but otherwise acts as the identity.

As another variation on~\eqref{eq:CCZ} we can represent the {CCNOT} gate as follows:
\begin{equation}
%
\end{equation}

Those familiar with the ZX-calculus or the ZH-calculus might have noticed that they have conflicting definitions of the X-spider and the 2-ary H-box, resulting in different scalar factors of $\sqrt{2}$. In this paper we use the conventions also used in \textsc{PyZX}~\cite{kissinger2019Pyzx} in order to aid in our calculations.
This means that our Z- and X-spider are defined as is usual in the ZX-calculus. However, most literature on the ZX-calculus also includes a yellow box to represent the Hadamard gate. In our case we use the convention of the ZH-calculus that such a box represents an \emph{unnormalised} Hadamard gate (cf.~\eqref{eq:Hdef}). Hence, certain scalar factors will be different than is usual in the literature on the ZX-calculus. Conversely, our H-box and Z-spider match the definition used in the ZH-calculus, but our X-spider does not match the corresponding definition in the ZH-calculus, and is off by certain factors of $\sqrt{2}$.
It is unfortunately not possible to have a fully satisfactory convention when it comes to scalar factors in the ZX/ZH-calculus, and choices have to be made about where scalar corrections appear (see~\cite{debeaudrap2020welltempered} for a longer discussion on this topic). 
In order to prevent confusion about these clashing scalar conventions, we will refer to our version of the ZX and ZH calculus as the \emph{ZXH-calculus} throughout the paper.

\subsection{Graph states}

As it will be important for Section~\ref{sec:2DAKLT}, let us recall briefly the notion of \emph{graph states} and how they can be represented in the ZX-calculus.
Given a simple undirected graph $G=(V,E)$, there is a corresponding graph state $\ket{G}$. The state $\ket{G}$ is constructed by preparing for each vertex $v\in V$ a qubit in the $\ket{+}$ state, and for each edge $(v_1,v_2)\in E$ applying a CZ gate between the qubits corresponding to $v_1$ and $v_2$~\cite{graphstates}. 
Recall that graph states are important as all stabiliser states can be reduced to a graph state (up to local Cliffords)~\cite{elliott2008graphical}, and because most measurement-based quantum computation protocols use a graph state as their resource state~\cite{MBQC1}.

The representation of a graph state in the ZX-calculus is most easily explained by an example:
\begin{equation}\label{eq:graph-state-example}
\begin{tikzpicture}
	\begin{pgfonlayer}{nodelayer}
		\node [style=vertex] (0) at (-2.25, 1.75) {};
		\node [style=vertex] (1) at (-2, 0) {};
		\node [style=vertex] (2) at (0.25, 1) {};
		\node [style=vertex] (3) at (-0.5, 2.5) {};
		\node [style=vertex] (4) at (1, -0.75) {};
		\node [style=none] (5) at (-0.75, -1.5) {$G$};
		\node [style=Green Node] (6) at (6.5, 1.75) {};
		\node [style=Green Node] (7) at (6.75, 0) {};
		\node [style=Green Node] (8) at (9, 1) {};
		\node [style=Green Node] (9) at (8.25, 2.5) {};
		\node [style=Green Node] (10) at (9.75, -0.75) {};
		\node [style=none] (11) at (8, -1.5) {$\ket{G}$};
		\node [style=none] (12) at (4, 1) {$\rightarrow$};
		\node [style=none] (13) at (10.75, 2.5) {};
		\node [style=none] (14) at (10.75, 1.75) {};
		\node [style=none] (15) at (10.75, 1) {};
		\node [style=none] (16) at (10.75, 0) {};
		\node [style=none] (17) at (10.75, -0.75) {};
		\node [style=H] (18) at (7.45, 2.175) {};
		\node [style=H] (19) at (9.175, -0.325) {};
		\node [style=H] (20) at (8.45, 0.725) {};
		\node [style=H] (21) at (8.075, 1.3) {};
		\node [style=H] (22) at (7.175, 0.7) {};
	\end{pgfonlayer}
	\begin{pgfonlayer}{edgelayer}
		\draw (0) to (3);
		\draw (0) to (2);
		\draw (2) to (4);
		\draw (0) to (4);
		\draw (1) to (3);
		\draw (1) to (2);
		\draw (6) to (9);
		\draw (6) to (8);
		\draw (8) to (10);
		\draw (6) to (10);
		\draw (7) to (9);
		\draw (7) to (8);
		\draw (10) to (17.center);
		\draw (7) to (16.center);
		\draw (8) to (15.center);
		\draw (6) to (14.center);
		\draw (9) to (13.center);
	\end{pgfonlayer}
\end{tikzpicture}%
\end{equation}
In words: for each vertex of the graph we add a Z-spider with a single output, and for each edge we add a corresponding wire between spiders with a Hadamard gate on it.

\section{The 1D AKLT state in the ZXH-calculus}\label{sec:AKLT-in-ZXH}

{\subsection{ZXH representation and relation to matrix-product states}}

We now have all we need to show how the AKLT state is represented in the ZXH-calculus. We start by representing the singlet operator $\ket{01}-\ket{10}$ of Fig.~\ref{fig:AKLT-mps}(a). 
Note that the Bell state $\ket{00}+\ket{11}$ is related to the singlet state by application of a Pauli $Z$ and $X$ on one of its qubits. Hence, the operator in ZXH is:
\begin{equation}\label{eq:singlet}
	\scalebox{1.0}{%
\begin{tikzpicture}
	\begin{pgfonlayer}{nodelayer}
		\node [style=none] (0) at (0, 1.25) {};
		\node [style=none] (1) at (0, -1.25) {};
		\node [style=Z phase dot] (2) at (-1, 0.5) {$\pi$};
		\node [style=X phase dot] (3) at (-1, -0.5) {$\pi$};
	\end{pgfonlayer}
	\begin{pgfonlayer}{edgelayer}
		\draw [bend right=90, looseness=1.50] (0.center) to (1.center);
	\end{pgfonlayer}
\end{tikzpicture}%
} \ \ = \ \ \ket{01}-\ket{10}.
\end{equation}
Indeed, an empty curved wire (commonly referred to as a `cup' in the ZX-calculus literature) is the Bell state $\ket{00}+\ket{11}$. 
If we then apply a Z $\pi$-phase ($\ket{0}\bra{0}+e^{i\pi}\ket{1}\bra{1}$) to the first (upper) qubit we get $\ket{00}-\ket{11}$. Applying a NOT gate (an X $\pi$-phase) on the second (lower) qubit we then get $\ket{01}-\ket{10}$ as desired.

The next operator we need to represent is the symmetriser on two spin-1/2 spaces. We encode the spin-1 state $\ket{+1}$ as the paired spin-1/2 state $\ket{00}$, the spin-1 state $\ket{0}$ as $\frac{\ket{01}+\ket{10}}{\sqrt{2}}$  and $\ket{-1}$ as $\ket{11}$. 
This is a convenient basis for us, and indeed the projector operator in Fig.~\ref{fig:AKLT-mps}(b) acts as the identity on this basis. 
In fact, the operator of Fig.~\ref{fig:AKLT-mps}(b) only acts to project away the $\ket{01}-\ket{10}$ state, which reduces the basis $\{\ket{00},\frac{\ket{01}+\ket{10}}{\sqrt{2}},\frac{\ket{01}-\ket{10}}{\sqrt{2}},\ket{11}\}$   into a three-dimensional space with basis $\{\ket{00},\frac{\ket{01}+\ket{10}}{\sqrt{2}},\ket{11}$.
We can represent the projection operator as a ZXH-diagram as follows:{
\begin{equation}\label{eq:O-operator}
	\scalebox{1.0}{%
\begin{tikzpicture}
	\begin{pgfonlayer}{nodelayer}
		\node [style=none] (0) at (-7.25, -1.25) {};
		\node [style=none] (1) at (-7.25, 1.25) {};
		\node [style=none] (2) at (-1.25, -1.25) {};
		\node [style=none] (3) at (-1.25, 1.25) {};
		\node [style=Green Node] (4) at (-2.25, -1.25) {};
		\node [style=Green Node] (5) at (-6.25, -1.25) {};
		\node [style=Green Node] (6) at (-4.25, -1.25) {};
		\node [style=Green Node] (7) at (-5.25, 0) {};
		\node [style=Green Node] (8) at (-4.25, 1.25) {};
		\node [style=Red Node] (9) at (-2.25, 1.25) {};
		\node [style=Red Node] (10) at (-6.25, 1.25) {};
		\node [style=H] (11) at (-4.25, 0) {};
		\node [style=H] (12) at (-5.25, -1.25) {};
		\node [style=H] (13) at (-3.25, -1.25) {};
		\node [style=none] (14) at (-8.5, 0) {\large$\frac{1}{2}$};
		\node [style=none] (15) at (0, 0) {$=$};
		\node [style=none] (30) at (3.375, 0) {$\begin{pmatrix}
1 & 0 & 0 & 0 \\
0 & \frac{1}{2} & \frac{1}{2} & 0
\\
0 & \frac{1}{2} & \frac{1}{2} & 0 \\
0 & 0 & 0 & 1 \end{pmatrix}$};
	\end{pgfonlayer}
	\begin{pgfonlayer}{edgelayer}
		\draw (0.center) to (5);
		\draw (4) to (2.center);
		\draw (5) to (12);
		\draw (12) to (6);
		\draw (6) to (13);
		\draw (13) to (4);
		\draw (6) to (11);
		\draw (11) to (8);
		\draw (11) to (7);
		\draw (10) to (1.center);
		\draw (3.center) to (9);
		\draw (9) to (8);
		\draw (8) to (10);
		\draw (5) to (10);
		\draw (4) to (9);
	\end{pgfonlayer}
\end{tikzpicture}%
}.
\end{equation}}

Indeed, this can be shown by checking its action on each of the basis states in $\{\ket{00},\frac{\ket{01}+\ket{10}}{\sqrt{2}},\frac{\ket{01}-\ket{10}}{\sqrt{2}},\ket{11}\}$ or composing the matrices presented in~\eqref{diagram:cnot} and~\eqref{eq:zero-projector}. We leave this as an exercise for the reader. Note how this diagram is symmetric under interchange of the inputs and outputs (i.e.~under a horizontal flip), and hence we will generally not care about its orientation in our diagrams.
We will find a different diagram that implements the same operator in Section~\ref{sec:higher-spins} where we show how to construct the symmetrising projector for larger Hilbert spaces.

In Figure~\ref{fig:AKLT-mps} we summarise our construction of the one\nobreakdash-dimensional AKLT state as a ZXH-diagram. 
We show the diagrammatic representation of its constituents, the singlet (Fig.~\ref{fig:AKLT-mps}(a)) and the projector (Fig.~\ref{fig:AKLT-mps}(b)). 
The ZXH-diagram of the 1D AKLT state is obtained by joining these in a (periodic) chain, as shown in Fig.~\ref{fig:AKLT-mps}(f).
This diagram consists of repetitions of the same block which is built out the symmetriser projector~\eqref{eq:O-operator} (Fig.~\ref{fig:AKLT-mps}(b)) and singlets~\eqref{eq:singlet} (Fig.~\ref{fig:AKLT-mps}(a)): 
\begin{equation}
	\centering
	\scalebox{0.8}{%
%
}
	\label{eq:1D-aklt}
\end{equation}
{
Note how we have a Z $\pi$ phase on the left, respectively an X $\pi$ phase on the right, open wires. We are allowed to do this as we are free to choose a basis for these degrees of freedom. We choose this convention as it allows us to see that there is a repeating block structure.
}
We can now show explicitly how the ZXH-diagrammatic representation and the MPS representation of the AKLT state are connected. In Fig.~\ref{fig:AKLT-mps}(f) we have overlaid a gray box over the part of the ZXH-diagram that encodes the MPS matrices given in~\eqref{eq:AKLTMPS}, as we now show.

Recall that we represent the spin-1 $\ket{+1}$ state as $\ket{00}$ on a pair of spin-1/2 wires. If we apply this state, given by the first diagram in Eq.~\eqref{diagram:basis-demo}, to one of the sites of \eqref{eq:1D-aklt}, we get a diagram that can be drastically simplified and be shown to be equal as a matrix to $M^{[n]+1}$ up to a scalar factor of $\frac{1}{\sqrt{6}}$:

\begin{equation}\label{eq:000-reduction}
	\scalebox{1.0}{%
%
}
\end{equation}
As we are plugging $\ket{00}$ into the top wires, we start with a scalar $\frac12$ as shown in \eqref{diagram:basis-demo}.
Note that in the last diagrammatic step we used that a Z-spider with no legs is equal to a scalar $2$.
The reason we keep track of scalars here is because for the MPS representation it is important that the matrices are scaled correctly with respect to each other.

We now proceed analogously, showing that if we plug the two remaining spin\nobreakdash-1 states, $\ket{0}$ and $\ket{-1}$, into one of the sites of \eqref{eq:1D-aklt} that we get the corresponding MPS matrices up to the same scalar factor of $\frac{1}{\sqrt{6}}$. First, we obtain $M^{[n]0}$ by plugging $\frac{1}{\sqrt{2}}(\ket{01}+\ket{10})$, which corresponds to the $\ket{0}$ spin-1 state:
\begin{equation}\label{eq:order-reduce}
	\scalebox{1.0}{%
%
}
\end{equation}
And similarly, we obtain $M^{[n]-1}$:
\begin{equation}\label{eq:111-reduction}
	\centering
	\scalebox{1.0}{%
%
%
}
\end{equation}
Note here that the last instance of \CopyRule introduced an $e^{i\pi} = -1$ scalar.

As summarized in Fig.~\ref{fig:AKLT-mps}(e), Eqs.~\eqref{eq:000-reduction}, \eqref{eq:order-reduce} and \eqref{eq:111-reduction} show that the ZXH representation encodes the same information as the MPS representation Eq.~\eqref{eq:AKLTMPS}, up to a global factor that can be fixed by normalising the state. We can conclude that our ZXH-diagram is indeed equal to the AKLT state. The advantage of the ZXH representation is that we can compute with it diagrammatically, as we will now show.  

{\subsection{Edge states and string order}}

From the ZXH-diagram of the 1D AKLT state in Eq.~\eqref{eq:1D-aklt} and Fig.~\ref{fig:AKLT-mps}(f) we can immediately infer one of its main properties: the presence of spin-1/2 edge states under open boundary conditions. Observe that the finite chain Eq.~\eqref{eq:1D-aklt} has two dangling wires at the bottom on the left and on the right.
{ The precise way of ending the chain amounts to a choice in boundary conditions, as in a conventional MPS, which fixes the edge two-dimensional spin-1/2 degrees of freedom~\cite{Moudgalya2018}. If the boundary condition is not fixed, the dangling edge wires can be understood as the projective (or fractionalized) symmetry representation of the bulk spin-1 rotation symmetry~\cite{Pollmann:2012gb}.}

A second property of the AKLT state, the non-zero string order parameter, can be shown by direct computation on its ZXH-diagram as follows.
We take $L$ sites in a chain, and we post-select each of the physical indices on the sites $2,3,\ldots, L-1$ to the state $\ket{0}$:
\begin{equation}\label{Order-op-equation}
	\scalebox{0.83}{%
%
}
\end{equation}
The non-vanishing of the string order parameter then tells us that the sites $1$ and $L$ cannot then both be in the spin $+1$ or spin $-1$ state.
On the level of the diagram we can see this behaviour when we post-select both of the states $j_1$ and $j_L$ to the same non-zero spin state:
\begin{equation}
	\centering
	\label{eq:AKLTorder0}
	\scalebox{0.80}{%
%
%
}
\end{equation}
That this diagram is zero tells us that the spin configuration where $j_1$ and $j_L$ are equal is not part of the AKLT state.

In contrast, when $j_{1}\neq j_{L}$ we get
\begin{equation}
	\centering
	\label{eq:AKLTorder1}
	\scalebox{0.80}{%
%
%
}
\end{equation}
Hence, the configuration where $j_1\neq j_L$ is part of the AKLT state.
These results signify the dilute anti-ferromagnetic order characteristic of the 1D AKLT state. 

While one could use software such as the \textsc{PyZX Python} package~\cite{kissinger2019Pyzx} to simplify the diagrams above to show that these diagrams are indeed (non-)zero, it is illustrative to rewrite the diagram manually. 
Note that the central repeated building block consisting of the projection to the spin-1 subspace followed by a post-selection for the $\ket{0}$ spin-1 state is exactly the diagram we simplified in~\eqref{eq:order-reduce}. Hence, \eqref{Order-op-equation} simplifies to:
\begin{equation}
	\scalebox{0.85}{%
%
}
\end{equation}
Note that this diagram is only equal to \eqref{Order-op-equation} up to non-zero scalar, but as we only care about whether the coming diagrams are zero or not, this is enough for our purposes.
Depending on the number of repetitions of the central block this diagram simplifies to one of the following:
\begin{equation}\label{eq:order-reduce-simplified}
	\centering
	\scalebox{1.0}{%
%
%
}
\end{equation}
Whether this middle $Z$ phase appears depends on whether there are an even or odd number of intermediate $\ket{0}$ states applied - giving a Z $\pi$-phase in the former case and none in the latter.
Now suppose we take $j_1=j_L = \ket{+1}$. Then we get the following diagram and simplification:
\begin{equation}\label{eq:order-param-numbers-00}
	\scalebox{1.0}{%
%
%
}
\end{equation}
A spider with a phase $\pi$ with no legs is equal to $1+e^{i\pi} = 1-1 = 0$, and hence this is indeed zero as we expect.
The case where we take $j_1=j_L = \ket{-1}$ is shown similarly.
Now when we set $j_1\neq j_L$, for instance, $j_1=\ket{-1}$ and $j_L=\ket{+1}$ we get a non-zero diagram:
\begin{equation}\label{eq:order-param-numbers-10}
	\scalebox{1.0}{%
%
%
}
\end{equation}
Indeed, as the scalar red spider we get is equal to $2$, this diagram is indeed non-zero.

To summarise: we started with the 1D AKLT chain \eqref{eq:1D-aklt}. We then post-selected an arbitrary number of adjacent sites to the spin-1 $\ket{0}$ state, resulting in the diagram~\eqref{Order-op-equation} which we simplified to one of the diagrams in~\eqref{eq:order-reduce-simplified} depending on the parity of the number of $\ket{0}$ sites. Then, in Eq.~\eqref{eq:order-param-numbers-00} we saw that post-selecting the $j_1$ and $j_L$ sites to be equal but non-zero spins resulted in a zero diagram. However, in ~\eqref{eq:order-param-numbers-10} we saw that post-selecting the $j_1$ and $j_L$ sites to be different non-zero spins resulted in a non-zero diagram. These observations signal the non-vanishing of the anti-ferromagnetic string order, as expected for the AKLT state.

The calculations presented in this section are also available in the accompanying Jupyter notebook.\footnote{Click   \href{https://github.com/Quantomatic/pyzx/blob/4837ea92ec56a98af268401a2c3fcb32946d5faa/demos/AKLT/AKLT\%20chain\%20demonstration.ipynb}{here} to see the relevant Jupyter notebook.
}

{
\subsection{Quantized Berry phase}\label{sec:Berry-phase}

We now show how to calculate the Berry phase for the 1D AKLT state~\cite{Hatsugai:2006hb} diagrammatically, obtaining an exact result for any finite chain. To calculate the Berry phase one introduces a phase twist within a given bond (a phase in our case, but a unitary matrix in general). For the periodic 1D AKLT state, this amounts to picking one singlet of the AKLT state $\ket{\psi}$ and transforming it to  $\ket{10}-e^{i\theta}\ket{01}$. This defines a twisted AKLT state $\ket{\psi_{\theta}}$ for each angle $\theta$ and we recover the standard 1D AKLT state when $\theta=0$~\cite{Hatsugai:2006hb}. 
The Berry phase is then defined as
\begin{equation}
\label{eq:Berryphasedef}
    \gamma= -i \int_{0}^{2 \pi}\dfrac{\left\langle\psi_{\theta}\left|\partial_{\theta}\right| \psi_{\theta}\right\rangle}{\braket{\psi_\theta}{\psi_\theta}}d\theta,
\end{equation}
were we have used the expression for an unnormalised wavefunction (see e.g.\cite{Ji2020}) in terms of the normalisation factor $\braket{\psi_{\theta}}{\psi_{\theta}}$. 
}

{
To calculate this value diagrammatically we start by writing the twisted AKLT state $\ket{\psi_{\theta}}$ as a ZXH diagram:
\begin{equation}
	\scalebox{0.9}{%
%
}
\end{equation}
To obtain the Berry phase we need to take the derivative of this diagram. For this we could use the techniques for diagrammatic differentiation described in~\cite{zhao2021analyzing,toumi2021diagrammatic}. For our purposes however it suffices to derive a couple of simple equations which can then be described as diagrams. In particular we will need the following diagram equality:
\begin{equation}
	\scalebox{1}{%
%
%
}
\end{equation}
%
%
%
Here the factor of $\frac{1}{2}$ is introduced because single-wire spiders are equal to states up to a constant $\sqrt{2}$.
Using this identity we get:
\begin{equation}
\partial_{\theta}\ket{\psi_\theta}\ \ =\frac{ie^{i\theta}}{2}\ \ \scalebox{0.9}{%
%
%
}
\end{equation}
We then have the integrand of the unnormalised Berry phase over which we must integrate:
\begin{equation}\label{eq:berry-phase-integral}
\gamma =(-i)\int_{0}^{2 \pi} \frac{ie^{i\theta}}{2\braket{\psi_\theta}{\psi_\theta}}\  \ \scalebox{0.75}{%
%
%
}\ d\theta
\end{equation}
We can simplify this expression somewhat by combining the adjacent symmetrisers:
\begin{equation}\label{eq:berry-simplify}
	\scalebox{0.9}{%
%
%
}
\end{equation}
Now, in order to calculate the expression of~\eqref{eq:berry-phase-integral} we split the diagram there up into two terms, using the following identity:
\begin{equation}\label{eq:phase-decomp}
 \scalebox{0.95}{%
%
%
}
\end{equation}
So, using~\eqref{eq:berry-simplify} and~\eqref{eq:phase-decomp} in~\eqref{eq:berry-phase-integral} we arrive at:
\begin{equation}
\gamma = \int_{0}^{2 \pi}\!\!\frac{2^{N-2}}{\braket{\psi_\theta}{\psi_\theta}}\left( e^{i\theta}\ \scalebox{0.47}{%
%
%
} \right) d\theta \label{EQ:Berry-integrand}
\end{equation}
Here $N$ is the length of the chain and the $2^N$ term comes from repeated application of~\eqref{eq:berry-simplify}.

To arrive at an equation for all $N$ we must decompose our diagrams in some systematic scalable fashion. To this end we use the following identities (see Appendix~\ref{app:Berry-phase-elements} for the proofs):
\begin{equation}
\label{eq:calcone}
	\scalebox{0.9}{%
%
%
}
\end{equation}
We can similarly derive:
\begin{equation}
\label{eq:calcfour}
	\scalebox{0.9}{%
%
%
}
\end{equation}
See~\eqref{equation:Berry-chain-norm} for the proof (and note that they are mirror images of each other). Now let's simplify the first term in the integrand of~\eqref{EQ:Berry-integrand}. We do this by repeatedly applying~\eqref{eq:calcfour}:
\begin{equation}\label{eq:integrand1-reduce}
	\scalebox{0.6}{%
%
%
}
\end{equation}
Hence, this is equal to: $4(-1)^{N}$.
We can similarly simplify the second term in the integrand by using Eqs.~\eqref{eq:calcone}, \eqref{eq:calctwo} and~\eqref{eq:calcthree}:
%
%

%

\begin{equation}
\label{eq:integrand2-reduce}
	\scalebox{0.7}{%
%
%
}
\end{equation}
So the value is:
\begin{equation}\label{eq:interm}
	\scalebox{0.9}{%
%
%
}
\end{equation}
We can fuse all the red spiders and then, using the equality $\ket{0} = \frac{1}{\sqrt{2}}(\ket{+}+\ket{-})$, we can reduce it to a sum of two simpler diagrams: 
\begin{equation}
	\scalebox{0.8}{%
%
%
}
\end{equation}
Here $a\in\{0,1\}$ depends on whether $N$ is even ($a=0$) or odd ($a=1)$, and we write $c=\left(\frac{1}{\sqrt{2}}\right)^{N-2} = \frac{2}{\sqrt{2}^N}$.

Now that we know the value of the two terms of the integrand, it remains to calculate the normalisation factor $\braket{\psi_\theta}{\psi_\theta}$. We first simplify the diagram by combining symmetrisers using~\eqref{eq:berry-simplify} and then decompose the $\theta$-labelled spiders using~\eqref{eq:phase-decomp} twice to get the normalisation factor:
\begin{equation}\label{eq:norm-decomp}
	\scalebox{0.75}{%
%
}
\end{equation}

Each of the four diagrams we get on the right we have already calculated the value of. The diagrams for $a=b=0$ and $a=b=1$ are equal to that in~\eqref{eq:integrand2-reduce}, while the other two are equal to those in~\eqref{eq:integrand1-reduce}. Hence:
\begin{align}\label{eq:norm}
	\ketbra{\psi_\theta}{\psi_\theta} &= \frac14 2^N ((e^{i\theta}+e^{-i\theta})\cdot 4(-1)^N +  2\cdot 2(3^N+(-1)^N)) \nonumber \\
	&= 2^{N}(2(-1)^N\cos\theta  + 3^N+(-1)^N)
\end{align}
%
%
%
It is simple to check that for $\theta=0$ the norm can be rewritten as $6^N + 3(-2)^{N}=(\sqrt{6})^{2N}(1+3(-1/3)^{N})$, which coincides with the usual AKLT normalisation (see e.g. below equation (90) in \cite{Schollwock11}) up to the prefactor $(\sqrt{6})^{2N}$. 
This different prefactor is the same $\sqrt{6}$ factor as seen in Eqs.~\eqref{eq:000-reduction},\eqref{eq:order-reduce} and \eqref{eq:111-reduction}.


Combining Eqs.~\eqref{eq:norm}, \eqref{eq:integrand1-reduce} and~\eqref{eq:integrand2-reduce} the Berry phase is given by
\begin{align}
	\gamma &= \int_{0}^{2\pi}\frac{2^{N-2}\left(4(-1)^Ne^{i\theta} + 3^N + (-1)^N\right)}{2^N\left(2(-1)^N\cos\theta + 3^N + (-1)^N\right)}d\theta \\
	&=\frac12 \int_0^{2\pi} \frac{2(-1)^N e^{i\theta} + 3^N + (-1)^N}{2(-1)^N\cos\theta + 3^N + (-1)^N}d \theta
\end{align}
Now, factor out the term $3^N + (-1)^{N}$ from the fraction and define the constant
\begin{equation}
\label{eq:a}
g=\frac{2(-1)^N}{3^N+(-1)^N}.
\end{equation}
We then see that
\begin{eqnarray}
\nonumber
\gamma &=&\frac{1}{2}\int_{0}^{2 \pi}\frac{g e^{i\theta}+1}{g\cos\theta+1} d\theta =
\frac{1}{2}\int_{0}^{2 \pi}\frac{g \cos\theta+ ig\sin \theta + 1}{g\cos\theta + 1} d\theta \\
 &=& \frac{1}{2}\left(\int_{0}^{2 \pi} 1 d\theta + \int_{0}^{2 \pi}\frac{ig\sin \theta}{(1+g\cos\theta)} d\theta\right) \nonumber \\
 &=& \frac12 (2\pi + 0) = \pi
\end{eqnarray}
Here the second integral evaluates to zero because it is an odd function. 
%
We thus arrive to $\gamma = \pi$ as was already known in the thermodynamic limit, but which here is shown to hold for all finite lengths~\cite{Hatsugai:2006hb}. 
}

\section{Encoding higher spins in multiple wires}\label{sec:higher-spins}

The wires in a ZXH-diagram represent two-dimensional Hilbert spaces, or in other words, they carry a spin-1/2 representation of $SU(2)$.
In the previous section we represented a spin-1 wire (a three-dimensional Hilbert space) by a pair of spin-1/2 wires together with a projector to the appropriate subspace.
This raises the question of how we can generalise this construction to higher spin representations, and thus larger Hilbert spaces.

To do this we need some basic representation theory. 
Recall that the group $SU(2)$ has a unique irreducible representation on $\C^n$ for each $n$~\cite{hall2015lie}. For $n=1$ this is the trivial representation, and for $n=2$ this is the fundamental representation where each matrix $M$ simply acts by matrix multiplication. 
For our purposes a convenient way to write the $n$-dimensional irreducible representation of $SU(2)$ (which is spin-$n/2$) is to take the \emph{symmetric subspace} of $n$ copies of the fundamental representation~\cite{martin2019primer}.  That is, we build spin-$n/2$ from the \emph{symmetric subspace} of $n$ spin-1/2 spaces. So, starting with the space of the fundamental representation $\mathcal{H} = \C^2$ we build the the space of the $(n+1)$-dimensional representation as $\text{Sym}(H^{\otimes n})$.
Indeed, $\text{Sym}(H^{\otimes n})$ has dimension $n+1$ as a basis for it is given by $\ket{0\cdots 0}$, $\ket{1\cdots 1}$ and the uniform superpositions of computational basis states containing a fixed number of $\ket{1}$'s, such as $\ket{10\cdots0}+\ket{010\cdots 0}+\cdots +\ket{0\cdots 01}$.
In summary, if we can represent the projector to the symmetric subspace on $n$ wires as a ZXH-diagram then we will have succeeded in representing arbitrary-dimensional (spin) systems on a collection of qubit wires. 

Let $\sigma\in S_n$ be a permutation on $n$ points. We write $U_\sigma$ for the unitary on $\mathcal{H}^{\otimes n}$ that permutes each of the composite $\mathcal{H}$ spaces via $\sigma$: $U_\sigma\ket{x_1\cdots x_n} = \ket{x_{\sigma 1}\cdots x_{\sigma n}}$.
Note that a space can be symmetrised by taking the superposition over all the permutations. Hence, the \emph{symmetrising projector} $P_S^{(n)}$ on $n$ wires is given by
\begin{equation}
    P_S^{(n)} := \frac{1}{n!} \sum_{\sigma \in S_n} U_\sigma.
\end{equation}

Each $U_\sigma$ can straightforwardly be written as a ZXH-diagram by just permuting the wires, but as we need to represent a coherent superposition of these permutation unitaries we need a controlled permutation operator. It turns out to be sufficient to use \emph{controlled SWAP} (CSWAP) operators that have the
control qubit post-selected into $\bra{+}$.
Recall that the CSWAP is defined by $\ket{0xy} \mapsto \ket{0xy}$ and $\ket{1xy} \mapsto \ket{1yx}$, i.e.~the first qubit determines whether the second and third qubit are swapped.
Including the post-selection, we can represent this (up to a non-zero scalar) as a ZXH-diagram in a convenient way:
\begin{equation}\label{diagram:nice-swap-open}
	\scalebox{1.0}{%
%
}
\end{equation}
We will refer to the right-hand side as a CSWAP in what follows.

Here the top wire is the control. By inputting a computational basis state we can verify that it indeed performs the maps required.
First, when the input is $\ket{0}$:
\begin{equation}
\scalebox{1.0}{%
%
%
}
\end{equation}
And second, when the input is $\ket{1}$:
\begin{equation}
\scalebox{1.0}{%
%
%
}
\end{equation}

Now, to write the symmetrising projector on $n=2$ wires we need an equal superposition of the identity permutation and the SWAP. Hence, if we make the control of
\eqref{diagram:nice-swap-open} a $\ket{+} = \ket{0}+\ket{1}$ state we get the desired map:
\begin{equation}\label{eq:CSWAP-superposition}
%
%
\end{equation}

To generalise this to larger $n$ we use induction. Indeed, if we have a coherent superposition of all the permutations on $n$ wires $P_S^{(n)}$, then to get a coherent superposition of the permutations on $n+1$ wires we need to compose $P_S^{(n)}$ with a coherent superposition of the identity and the SWAP gates from the $(n+1)$th qubit to every other qubit: $\text{id}+\text{SWAP}_{1,n+1}+\text{SWAP}_{2,n+1}+\cdots+\text{SWAP}_{n,n+1}$.
We construct this superposition as a ZXH-diagram by writing CSWAP gates from the $(n+1)$th qubit to each other qubit and then connecting all the control wires in such a way that at most one CSWAP `fires' at the same time. This gives us the general construction for $n$ wires.

For $n=3$ this gives the following diagram:
\begin{equation}
	\centering
	\scalebox{1.0}{%
%
}
	\label{diagram:3-symmetriser}
\end{equation}
This works, because
\begin{equation}
%
%
 \ \ \propto \ \ \ket{00} + \ket{10} + \ket{01},
\end{equation}
and hence the appropriate superposition is created.

{
We present a general construction for higher spins in Appendix~\ref{app:higher-spins}.
}

Notice that the symmetric subspace encoding for two wires of {\eqref{eq:CSWAP-superposition}} seems to give an alternative form of the symmetrising projection given in \eqref{eq:O-operator}. They can however be shown to be equal, up to an irrelevant scalar:
\begin{equation}\label{eq:1D-aklt2}
	\scalebox{0.85}{%
%
%
}
\end{equation}
and as such our 1D AKLT chain  (cf.~Eq.~\eqref{eq:1D-aklt}) can alternatively be written as:
\begin{equation}\label{eq:1D-aklt3}
	\scalebox{1.0}{%
%
%
}
\end{equation}
Where the projector now is of the form \eqref{eq:CSWAP-superposition}.

Note that there are modified versions of the ZX-calculus where a wire carries a three-dimensional Hilbert space~\cite{Qtritdit,EPTCS266.3}. However, much less is known about rewriting those diagrams, and it is harder to reason about the types of diagrams we have in this paper where we mix systems of different types of spins.

\section{The 2D AKLT state as a universal resource for quantum computing}\label{sec:2DAKLT}
\begin{figure*}
\center{
\hspace{-18.5cm} 
	\includegraphics[width=1.0\textwidth]{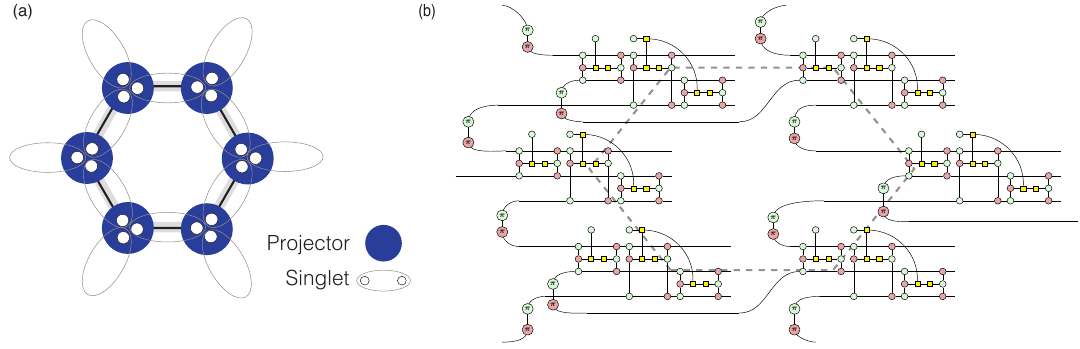}}
	\caption{The 2D AKLT state on a hexagonal lattice and its representation as a ZXH-diagram. (a) Pictorial representation of the unit cell of the 2D AKLT state on a hexagonal lattice. At each site there is a spin-3/2 degree of freedom that can be decomposed into three spin-1/2 states that form singlets with their nearest neighbours (represented by oval shapes). The blue circles denote projectors to the appropriate symmetric subspace. The gray hexagon denotes a choice of unit cell. (b) The 2D AKLT state unit cell as a ZXH-diagram, with the same unit cell denoted by a gray dotted line.}
	\label{fig:2DAKLT}
\end{figure*}
We will now study the generalization of the 1D AKLT state to the 2D hexagonal lattice~\cite{AKLT87}, depicted in Fig.~\ref{fig:2DAKLT}(a). 
First, we derive the representation of this state as a ZXH-diagram, and then  we show how it can be used as a universal resource for quantum computing, by showing that it reduces to a graph state.

As mentioned in the introduction, it is possible to construct an AKLT type state on a hexagonal lattice using spin-3/2 degrees of freedom at each site (Fig.~\ref{fig:2DAKLT}(a)). Each spin-3/2 degree of freedom corresponds to a four-dimensional Hilbert space and, by the discussion in the previous section, can be represented on a set of three qubit wires with the projector presented in \eqref{diagram:3-symmetriser}.
So whereas in the 1D AKLT state we projected two spin-1/2 states down to the symmetric subspace to represent a spin-1 degree of freedom, here we project three spin-1/2 degrees of freedom to form a spin-3/2.
This projector, with each of the component spin-1/2 wires linked to another by singlet states, forms the basic unit (a site) of the 2D AKLT state. As a ZXH-diagram:
\begin{equation}
	\centering
	\scalebox{1.0}{%
%
}
	\label{eq:ZXHspin32}
\end{equation}
Here we have a single spin-3/2 degree of freedom of the 2D AKLT state with singlet states on each of its legs. These can then be combined to give a diagram of a lattice that is not just a convenient visual aid for the 2D AKLT state, but literally \emph{is} the 2D AKLT state; see Figure~\ref{fig:2DAKLT}(b).

Analogous to the 1D AKLT example in Fig.~\ref{fig:AKLT-mps} where two wires corresponded to the physical spin-1 state, the triples of wires coming out to the right of \eqref{eq:ZXHspin32} correspond to the physical spin-3/2 degrees of freedom that form the state. The remaining wires of the diagram should be considered to be connected to other parts in the hexagonal lattice periodically (see Fig.~\ref{fig:2DAKLT}(b)).\\

We will now show how a hexagonal lattice AKLT state reduces to a graph state under a suitable measurement of the spin-3/2 degrees of freedom.
A consequence of this result is that the 2D AKLT state is a universal resource for measurement-based quantum computing~\cite{van2006universal}. This result was already shown in Ref.~\cite{wei2011affleck} and independently in Ref.~\cite{miyake2011quantum}. 
The proof in Ref.~\cite{wei2011affleck} consists of two parts. First, they showed the hexagonal lattice reduces to a graph state. Second, they used a percolation argument to prove the resulting state is a universal resource for quantum computation. We will derive the first part entirely diagrammatically.
In the process we will see that certain derivations concerning the simplification of the lattice presented in Ref.~\cite{wei2011affleck} are in our approach just the standard spider fusion rule \SpiderRule and the Hopf rule \HopfRule of the ZX-calculus.

To reduce the 2D AKLT state to a graph state, we need to reduce it to a simpler state. We do this by measuring each of the spin-3/2 states. Recall that each of these spin-3/2 states is presented as a symmetric three qubit state and hence a measurement on it can be present as a simultaneous measurement on these three qubits. The measurement is a POVM (Positive operator-valued measurement, the most general type of measurement~\cite{nielsen2002quantum}) with three elements:
\begin{align}
E_z &:= \frac23 (\ketbra{000}{000} + \ketbra{111}{111}), \\
E_x &:= \frac23 (\ketbra{+++}{+++} + \ketbra{---}{---}), \\
E_y &:= \frac23 (\ketbra{iii}{iii} + \ketbra{-i,-i,-i}{-i,-i,-i}).
\end{align}
Here the sets $\{\ket{0},\ket{1}\}$, $\{\ket{+},\ket{-}\}$ and $\{\ket{i},\ket{-i}\}$ denote respectively the eigenbases of the $Z$, $X$ and $Y$ Pauli matrices.
Usually the elements of a POVM should sum up to the identity, but as we are working in the symmetric subspace, we instead have $E_z+E_x+E_y = P_S$, where $P_S$ is the projection on the symmetric subspace, as desired.

Conveniently, each of these POVM elements can be represented as a small ZX-diagram (up to global scalar):
\begin{align}
E_z\ \  &\propto\ \  %
%
\label{eq:povm-y-3}
\end{align}
The forms of $E_z$ and $E_x$ follow directly from the definition of the Z- and X-spider.
To see the correctness of $E_y$ note that a Z $\frac{\pi}{2}$-rotation $R_z(\frac{\pi}{2})$ acts as $R_z(\frac{\pi}{2})\ket{+}= \ket{i}$ and $R_z(\frac{\pi}{2})\ket{-}= \ket{-i}$ where $\ket{\pm i}= \ket{0}\pm i\ket{1}$.
Hence, we can see~\eqref{eq:povm-y-3} as an X-projector surrounded by a basis transformation from the $Y$ eigenbasis to the $X$ eigenbasis. We could have equivalently chosen a Z-projector surrounded by X $\pm\frac{\pi}{2}$ rotations which corresponds to flipping the colours and the signs of the rotations; cf.~\cite[Section 9.4]{CKbook}. 
Note that $E_y$ is not symmetric under interchange of inputs and outputs, and thus unlike the case for $E_z$ or $E_x$, when considering $E_y$ we must keep in mind what we consider an input and output.

Importantly, each of the POVM elements $E_z$, $E_x$, $E_y$ projects to a 2D subspace, and hence encodes a spin-1/2 degree of freedom. While we could continue to work with the three output wires as a single qubit with the qubit operations encoded onto the three wires, we will instead represent the collapse to a single spin-1/2 degree of freedom by simply writing one wire:
\begin{align}
E_z\ \  &\rightsquigarrow\ \  %
\begin{tikzpicture}
	\begin{pgfonlayer}{nodelayer}
		\node [style=none] (27) at (-1.75, 1) {};
		\node [style=none] (28) at (-1.75, 0) {};
		\node [style=none] (29) at (-1.75, -1) {};
		\node [style=none] (49) at (1.75, 0) {};
		\node [style=Green Node] (51) at (0, 0) {};
	\end{pgfonlayer}
	\begin{pgfonlayer}{edgelayer}
		\draw (51) to (49.center);
		\draw [bend right, looseness=0.75] (51) to (27.center);
		\draw (51) to (28.center);
		\draw [bend left] (51) to (29.center);
	\end{pgfonlayer}
\end{tikzpicture}%
\label{eq:povm-z} \\
E_x\ \  &\rightsquigarrow\ \ %
\begin{tikzpicture}
	\begin{pgfonlayer}{nodelayer}
		\node [style=none] (52) at (-1.75, 1) {};
		\node [style=none] (53) at (-1.75, 0) {};
		\node [style=none] (54) at (-1.75, -1) {};
		\node [style=none] (58) at (1.75, 0) {};
		\node [style=Red Node] (60) at (0, 0) {};
	\end{pgfonlayer}
	\begin{pgfonlayer}{edgelayer}
		\draw (60) to (58.center);
		\draw [bend right, looseness=0.75] (60) to (52.center);
		\draw (60) to (53.center);
		\draw [bend left] (60) to (54.center);
	\end{pgfonlayer}
\end{tikzpicture}%
\label{eq:povm-x} \\
E_y\ \  &\rightsquigarrow\ \  %
\begin{tikzpicture}
	\begin{pgfonlayer}{nodelayer}
		\node [style=none] (61) at (-1.75, 1) {};
		\node [style=none] (62) at (-1.75, 0) {};
		\node [style=none] (63) at (-1.75, -1) {};
		\node [style=none] (67) at (1.75, 0) {};
		\node [style=Red Node] (69) at (0, 0) {};
		\node [style=Z phase dot] (70) at (-1, 0.95) {\ -$\frac\pi2\ $};
		\node [style=Z phase dot] (71) at (-1, 0) {\ -$\frac\pi2\ $};
		\node [style=Z phase dot] (72) at (-1, -1.025) {\ -$\frac\pi2\ $};
		\node [style=Z phase dot] (74) at (1, 0) {$\frac\pi2$};
	\end{pgfonlayer}
	\begin{pgfonlayer}{edgelayer}
		\draw (69) to (67.center);
		\draw [bend right, looseness=0.75] (69) to (61.center);
		\draw (69) to (62.center);
		\draw [bend left] (69) to (63.center);
	\end{pgfonlayer}
\end{tikzpicture}%
 \label{eq:povm-y}
\end{align}
We will use this `squigly arrow' $\rightsquigarrow$ to denote when we make a step that corresponds to a redefinition of the output basis. Here this is a collapse of a two-dimensional degree of freedom spread out over three wires to a single wire, but later on we will also use redefinitions to absorb single qubit gates that appear on output wires. Physically, this corresponds to updating the correspondence between the `logical' or `encoded' $\ket{0}$ and $\ket{1}$, and the actual physical states.

As these POVM elements are symmetric on the three qubits, they are preserved by the projection to the symmetric subspace, a fact we can prove diagrammatically.
For instance, considering $E_x$, we first show that it absorbs a CSWAP gate:
\begin{equation}\label{eq:x-povm-simp}
	\scalebox{0.8}{%
%
}
\end{equation}

Iterating this three times we then get the following equality:
\begin{equation}\label{eq:red-projector-alt}
	\scalebox{0.85}{%
%
%
}
\end{equation}
The floating scalar diagram on top multiplied by the scalar produced by the sequence of rewrites represents the eigenvalue of this operation under the projection.  This scalar is not important for our purposes, and we will drop it implicitly in later diagrams.

We can do a similar derivation for $E_z$ (see Appendix~\ref{app:povm-calc}):
\begin{equation}\label{equation:E_z_collapse}
	\scalebox{0.85}{%
%
%
}
\end{equation}
An analogous equation and derivation exists for $E_y$ as well (see Appendix ~\ref{app:povm-calc}).

We started with the 2D AKLT state on a hexagonal lattice (Figure~\ref{fig:2DAKLT}), and then we measured each of the spin-3/2 states  with this POVM $\{E_z,E_x,E_y\}$. 
Due to equations~\eqref{eq:red-projector-alt} and~\eqref{equation:E_z_collapse} and the analogous one for $E_y$, we see that regardless of the measurement outcome $E_z$, $E_x$ or $E_y$ that the symmetrising projector on each spin-3/2 output is `consumed' and replaced by the spider associated to one of $E_z$, $E_x$ and $E_y$.
Hence, what remains of the 2D AKLT state is a set of singlet states, connected via a network of spiders of the form~\eqref{eq:povm-z}--\eqref{eq:povm-y}. 
The state resulting from applying this measurement to the 2D AKLT state will hence be a hexagonal lattice where at each site we randomly have a X,Y or Z spider (which depends on the measurement outcome), and these are connected via singlet states. 
For example, the hexagonal unit cell of Fig.~\ref{fig:2DAKLT}(b) could be reduced to a diagram like the following:
\begin{equation}
	\centerline{
	\scalebox{0.8}{%
%
}}
\end{equation}
Readers familiar with the ZX-calculus can easily see that the resulting diagram is a \emph{Clifford diagram}. Indeed, it does not contain any higher-arity H-boxes, and the only phases that appear are multiples of $\frac\pi2$ making it a ZX-diagram in the Clifford fragment~\cite{Backens1}. As it only has outputs, it is a state, and hence is a Clifford state\footnote{Recall that a Clifford state, also called a \emph{stabiliser state}, is a state that is uniquely determined by being a eigenvalue 1 eigenvector of a set of Pauli operators. Any Clifford state can be represented by a ZX-diagram containing only spiders with phases that are multiples of $\frac{\pi}2$.}.
Any Clifford state can be presented as a graph state with single-qubit Clifford unitaries on its outputs (see Ref.~\cite{elliott2008graphical}, or for a proof using the ZX-calculus, see for instance Refs.~\cite{Backens1,duncan2019graph}).
Hence, we can already conclude that the state we get is a graph state.

\begin{figure*}
\hspace{-16.0cm}
\centering
\includegraphics[width=0.8\textwidth]{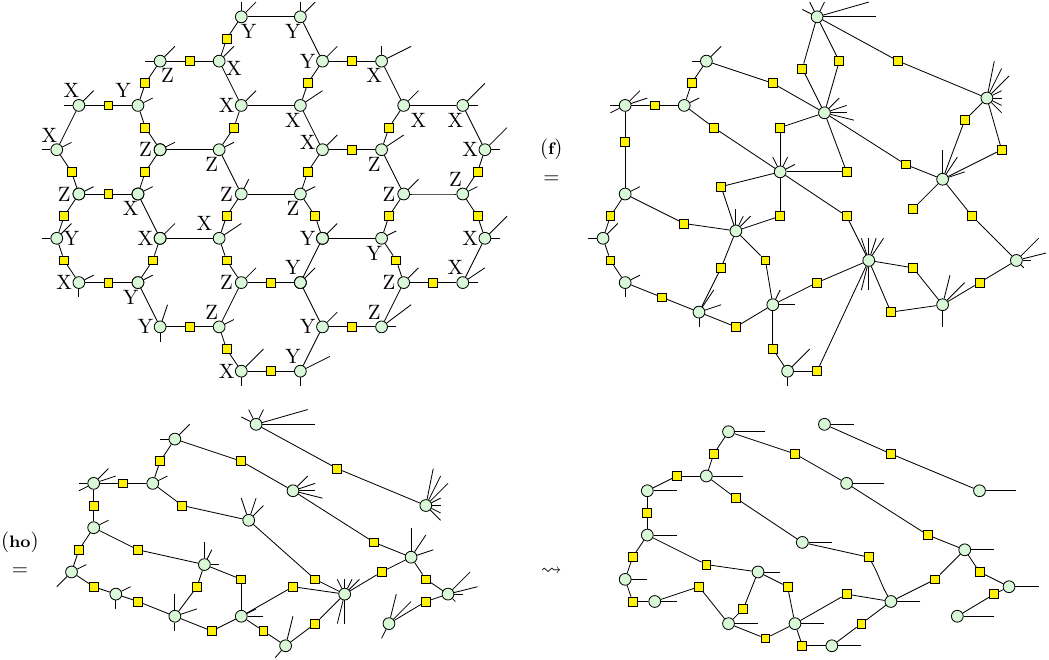}
\caption{ This figure shows the AKLT hexagonal lattice \emph{after} the $E_x$, $E_y$, and $E_z$ projectors have been applied and the $\pi$ phases moved onto the external wires and absorbed into basis redefinitions. 
In the first diagram on the left note that we have added X,Y,Z labels to the Z-spiders. These aren't formally part of the diagram, but are just labels to indicate which projector was applied to reach this diagram. The first equality shows that spiders with the same measurement outcome are merged.
Following this, the Hopf rule \HopfRule is applied to remove pairs of wires with a Hadamard box on them between the same spiders. 
The final step is to redefine the output basis to collapse multiple output wires coming from the same spider into a single wire. The resulting diagram can indeed be seen to be a graph state. This bears strong resemblance to the diagram seen in \cite{wei2011affleck} (figure 4 diagram C) where now their ad-hoc reduction is describable entirely in quantum-informational terms via the ZXH-calculus.
}
\label{fig:graph-red}
\end{figure*}

However, to show that the state we obtain is a universal resource for quantum computing we need to know more about the specific construction of the graph state, so let us go through the derivation manually. This happens in a few steps.

The first step is to get rid of the Z and X $\pi$-phases arising from the singlet states. 
We will do this by commuting these phases through the spiders onto the outputs of the state (the spin-3/2 outputs). 
For instance, for a $E_z$ outcome, we can do the following:
\begin{equation}
	\centering
	\scalebox{1.0}{%
\begin{tikzpicture}
	\begin{pgfonlayer}{nodelayer}
		\node [style=Red Node] (121) at (-3.75, 0.25) {$\pi$};
		\node [style=Green Node] (122) at (-2.5, 0.5) {$\pi$};
		\node [style=Green Node] (123) at (-6.75, 0) {$\pi$};
		\node [style=Red Node] (124) at (-5.75, 0) {$\pi$};
		\node [style=Green Node] (126) at (-4.75, 0) {};
		\node [style=none] (133) at (-4.75, 2) {};
		\node [style=none] (136) at (-7.75, 0) {};
		\node [style=Red Node] (141) at (-3.5, -0.5) {$\pi$};
		\node [style=Green Node] (142) at (-2.75, -0.75) {$\pi$};
		\node [style=none] (143) at (-1.5, -1.25) {};
		\node [style=none] (144) at (-1.5, 0.75) {};
		\node [style=none] (146) at (0, 0) {$=$};
		\node [style=Green Node] (169) at (3.75, 0) {};
		\node [style=Red Node] (171) at (3.75, 0.75) {$\pi$};
		\node [style=none] (173) at (1.5, 0) {};
		\node [style=none] (179) at (7, -1.25) {};
		\node [style=none] (180) at (7, 1) {};
		\node [style=Green Node] (181) at (3.75, 1.5) {$\pi$};
		\node [style=none] (182) at (3.75, 2) {};
		\node [style=none] (220) at (0, 1) {\spiderrule};
		\node [style=none] (221) at (0, 2) {\picopyrule};
	\end{pgfonlayer}
	\begin{pgfonlayer}{edgelayer}
		\draw (123) to (124);
		\draw (121) to (122);
		\draw (123) to (136.center);
		\draw (141) to (142);
		\draw (143.center) to (142);
		\draw (144.center) to (122);
		\draw (126) to (133.center);
		\draw (141) to (126);
		\draw (121) to (126);
		\draw (126) to (124);
		\draw (169) to (171);
		\draw (181) to (171);
		\draw (181) to (182.center);
		\draw (173.center) to (169);
		\draw (169) to (180.center);
		\draw (169) to (179.center);
	\end{pgfonlayer}
\end{tikzpicture}%
}
\end{equation}
Here the site is understood to be in the bulk of the lattice, with the top wire corresponding to its spin-3/2 degree of freedom\footnote{For sites that aren't in the bulk of the lattice, the calculation would be slightly different in that phases would pass onto the other external disconnected edges. However, these $\pi$ phases can be removed by redefining the basis of the external wires.}. 
Hence, we can remove the internal $\pi$ phases by moving them onto the external edges.
The analogous procedure for $E_x$ and $E_y$ measurement outcomes is demonstrated in Appendix~\ref{app:pi-extraction}.

Since each Z and X $\pi$-phase is connected to two spiders we need to make a choice about which way to commute each $\pi$. 
As the hexagonal lattice is two-colourable this is indeed possible in a consistent way.

After this procedure, we will have a diagram where the only $\pi$ phases are on the spin-3/2 outputs of the states.
As discussed beneath~\eqref{eq:povm-y}, our choice of representation of the spin-3/2 degree of freedom can be chosen arbitrarily. Hence we can redefine our basis here to remove these $\pi$ phases (this again corresponds to a redefinition of how we encode the $\ket{0}$ and $\ket{1}$ states on our physical system):
\begin{equation}
%
\end{equation}

The second step is to bring the diagram closer to the form of a graph state as presented in~\eqref{eq:graph-state-example} by changing the X-spiders coming from $E_x$ and $E_y$ measurement outcomes to Z-spiders.
This can be done easily using~\eqref{eq:X-to-Z}, and a redefinition of the output basis to remove the resulting Hadamard:
\begin{equation}
	\centering
	\scalebox{0.9}{%
%
%
}
\end{equation}
For the $E_y$ outcomes, we additionally remove the $\frac\pi2$ phases. For instance:
\begin{equation}
	\centering
	\scalebox{0.9}{%
%
%
}
\end{equation}
We leave the other cases to the reader.
The diagram we have now consists solely of Z-spiders and Hadamards.

Now, the third step of our reduction to a graph state is to fuse all the spiders that can be fused. In practice this means that two adjacent sites that had the same measurement outcome will be fused together. This fusing results in sites that have multiple outputs, which we again collapse to a single output as we did in~\eqref{eq:povm-z}--\eqref{eq:povm-y}. See Figure~\ref{fig:graph-red} for a demonstration of this procedure.

The final step is to remove parallel Hadamard-edges that could have been introduced by sites that were fused together. To do this we use a variation on the Hopf rule~\HopfRule:
\begin{equation}\label{eq:hopf-ZZ}
	\scalebox{0.9}{%
%
}
\end{equation}
The resulting diagram consists of phaseless Z-spiders connected via single Hadamard-edges, and hence is a graph state, as was desired.
Note that this entire procedure can also be done in an automated fashion using \textsc{PyZX}~\cite{kissinger2019Pyzx}; see the accompanying Jupyter notebook.\footnote{
Click \href{https://github.com/Quantomatic/pyzx/blob/4837ea92ec56a98af268401a2c3fcb32946d5faa/demos/AKLT/AKLT\%20hexagonal\%20lattice.ipynb}{here} to see the relevant Jupyter notebook.
}

Because neighbouring sites that have the same measurement outcome get fused, and parallel edges resulting from this fusing get disconnected, the highly regular hexagonal graph will generally collapse to a much less regular and more sparsely connected graph.
For example, consider the hexagonal graph given in Figure~\ref{fig:graph-red} where the vertices are labelled by $X$,$Y$, or $Z$ to denote the 2D AKLT state with the $E_x$, $E_y$ or $E_z$ measurement outcomes, and consider also its reduction with the rules outlined above.

Not any graph state can be used as a universal resource for measurement-based quantum computing.
The most canonical example of a universal resource state is the \emph{cluster state} that as a graph is just a regular square tiling.
In Ref.~\cite{wei2011affleck} it is shown via a percolation argument that given a large enough initial hexagonal lattice the irregular graph state resulting from the measurement of a 2D AKLT state can, with high probability, be further reduced to a cluster state. In particular, they show that the expected connectivity of the graph is above the critical `percolation threshold'~\cite{browne2008phase} which means that it includes a large cluster state subgraph with high probability.
Hence, for a large enough lattice we can use, with high probability, the 2D AKLT state to do universal measurement-based quantum computation.

{ 

\section{Crystal symmetries and transitions in ZXH}\label{section:symmeries}

\begin{figure*}
\hspace{-18.0cm} 
	\includegraphics[width=\textwidth]{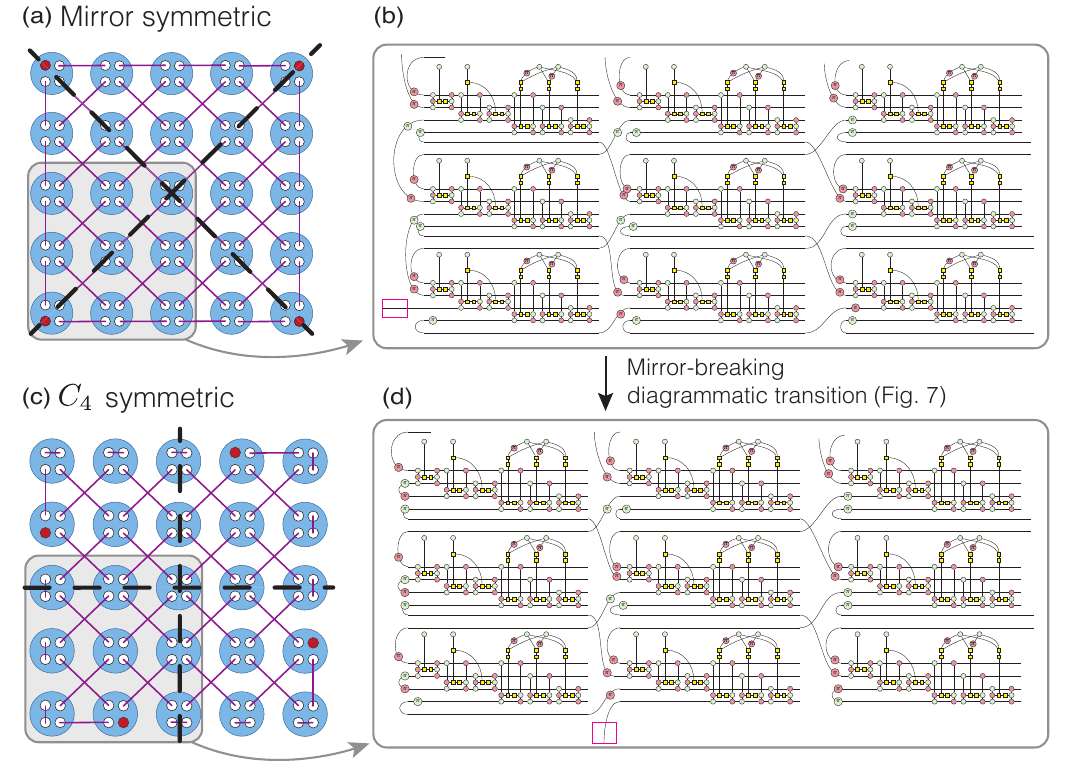}
	\caption{
	{ (a) A higher-order symmetry-protected topological phase with corner modes protected by mirror symmetries (diagonal dashed black lines).
	(b) The corresponding ZXH representation of the gray-shaded bottom-left quadrant.
	(c) When mirror symmetry is broken, $C_4$ symmetry protects the topological modes, which are then unpinned from the corners, as shown schematically. The transition between a higher-order topological state protected by mirror symmetry to one protected by $C_4$ symmetry can be modeled diagrammatically in ZXH, see Fig.~\ref{fig:sym2D}, resulting in the ZXH diagram (d). In (b) and (d) the topological modes are dangling wires (marked by the magenta box), and correspond to the red dots in the gray shaded areas in (a) and (c), respectively.}
	}
	\label{fig:HOTISPT}
\end{figure*}

Symmetries are at the core of our understanding of topological phases as they enrich their classification and simplify the calculation of topological invariants~\cite{Chiu16,Pollmann2010,Schuch2011,Chen2011,senthil2015}. One remarkable consequence of crystal symmetries, like rotation or mirror symmetries, is that they can protect gapless topological states not only at the boundaries of insulators, but also at the boundary of a boundary. For example, a 2D (respectively 3D) insulator with insulator edges (resp.~surfaces) can display protected (resp.~hinge) corner modes. These phases, known as higher-order topological insulators~\cite{Benalcazar:2017ks,Benalcazar:2017cn,Song:2017eva,Schindler2017,Schindler:2018hl}, can only exist in the presence of crystal symmetries. 

The goal of this section is to diagrammatically represent a transition between topological states with different crystal symmetries. Using mirror and rotational symmetries as a specific example, we will first discuss how to diagrammatically construct states that are symmetric crystal symmetries. This will require that the diagram representing the state is also symmetric, in a way that we will specify shortly. 
With these states in hand, we will construct a ZXH-diagram that transitions between two states with different crystal symmetries as a function of a control parameter. The possibility of diagrammatic transitions between topological states serves as an example of the potential of diagrammatic reasoning compared to other tensor networks, even for relatively simple states. 

Concretely, we consider the higher-order symmetry protected topological state based on the AKLT state shown in Fig.~\ref{fig:HOTISPT}(a)~\cite{Song:2017eva}.
Each site represents a spin-2 degree of freedom, which can be decomposed into four spin-$1/2$ wires. Coupling these spin-$1/2$s with singlets in the configuration shown pictorially in Fig.~\ref{fig:HOTISPT}(a) results in four unpaired spin-1/2 degrees of freedom that reside at the corners (red circles). The existence of each one of these unpaired spin-1/2 degrees of freedom is protected by mirror symmetry: they cannot be removed unless mirror symmetry is broken, for example by acting with different local unitary operators at sites related by mirror symmetry.  

Constructing this state as a ZXH-diagram is straightforward using our previous discussions. 
For each site we construct the $n=4$ symmetriser, as we did in Eq.~\eqref{diagram:3-symmetriser} for $n=3$. 
Then we connect the sites with singlets in the way specified in Fig.~\ref{fig:HOTISPT}(a). 
This results in the diagram shown in Fig.~\ref{fig:HOTISPT}(b), where we have only shown the lower-left quadrant for clarity. 
Note that we know that the symmetriser is symmetric under any permutation of its wires, by definition of it representing the symmetriser. 
Concretely this means it is irrelevant which intra-site wire connects to other sites as all wires within a site are equivalent. 
Hence, for the purposes of symmetry, any reordering of the connectivity at the individual site level is irrelevant and we need only to concern ourselves with the connectivity between different sites. 

So long as we connect sites in a way that respects the desired symmetry (which will be either mirror or rotational symmetry in our example) the diagram will possess the same crystal symmetries as the state it represents. 
This follows from a general property of ZXH-diagrams. 
If one constructs a diagram which can be brought to a symmetric form with respect to some lattice symmetry, the state that it represents must also have these symmetries. 
This is the case because the generating elements of the diagram, spiders and H-boxes, are themselves fully-symmetric tensors and thus any symmetry in the relation of the diagrammatic elements is \emph{also} a symmetry of the tensors they represent. Note this does not imply that an asymmetric diagram represents an asymmetric tensor, as it is is possible to apply rewrites to one side of a symmetric diagram to remove the diagrammatic symmetry. For instance, our symmetriser diagram is asymmetric, yet represents a symmetric tensor.
Note however that because the calculus is complete, there will always be a series of rewrites that transforms a diagram representing a tensor with some symmetry to a diagram with the same symmetry.

Instead of constructing a state with a mirror symmetry, as in Fig.~\ref{fig:HOTISPT}(a), we can similarly construct a state which has four-fold rotational symmetry; see Fig.~\ref{fig:HOTISPT}(c). This state also has dangling spin-$1/2$ states on each side, at positions related by $C_4$ symmetry.
Its corresponding ZXH representation is depicted in Fig.~\ref{fig:HOTISPT}(d), where once more we only show the lower-left quadrant for clarity. 

While a desirable property of ZXH-diagrams is that a symmetric diagram mathematically represents a symmetric state, one might feel that the schematic representations in Fig.~\ref{fig:HOTISPT}(a) and Fig.~\ref{fig:HOTISPT}(c) already imply that the states possess the symmetries we are interested in, even if they lack mathematical rigour. 
The ZXH-representations in and of themselves may then not seem like a sufficient advantage, at least for simple states. The advantage becomes clearer however when we consider what one can do once the states are rigorously defined. 
As we show next, the ZXH-diagrams allow us to go further than is possible with informal representations. We will show how to model a transition between these two states by diagrammatically breaking the symmetry. It is unclear how one would represent this schematically in a useful way. More importantly, it also goes beyond what one could achieve using other tensor network approaches, which would require explicit knowledge of the tensors that define these states.

Let us now describe how to interpolate between the mirror-symmetric state of Fig.~\ref{fig:HOTISPT}(a,b) and the $C_4$-symmetric state in Fig.~\ref{fig:HOTISPT}(c,d) using a parametrised ZXH-diagram.
Since we are dealing with (at least) $C_4$-symmetric states it is sufficient to focus on a quadrant, e.g. the bottom-left quadrant. Our goal will be to break the symmetry by moving the corner mode one site down, from the corner to the edge, along with the relevant inter-site singlets. 
We can represent the path between these two states by a parametrised ZXH diagram. 
To do this we will make repeated use of the following diagrammatic element that can represent both a singlet as well as a product state:
\begin{equation}\label{eq:entangler}
	\scalebox{1.0}{%
%
}
\end{equation}
When $f(\theta)=0$ it disconnects, while for $f(\theta)=\pi$ it generates a singlet between the spiders. These two cases are easily derived by application of \ExplodeRule and \AbsorbRule. 
By iterating this construction we can toggle the connectivity of many singlets at once in a diagram. It is precisely this mechanism that allows us to demonstrate a diagramatic transition between the two symmetric higher-order symmetry-protected states of Fig~\ref{fig:HOTISPT}; see Fig~\ref{fig:sym2D}.

\begin{figure*}
	\centering
	\hspace{-18.3cm} 
	\includegraphics[width=1.0\textwidth]{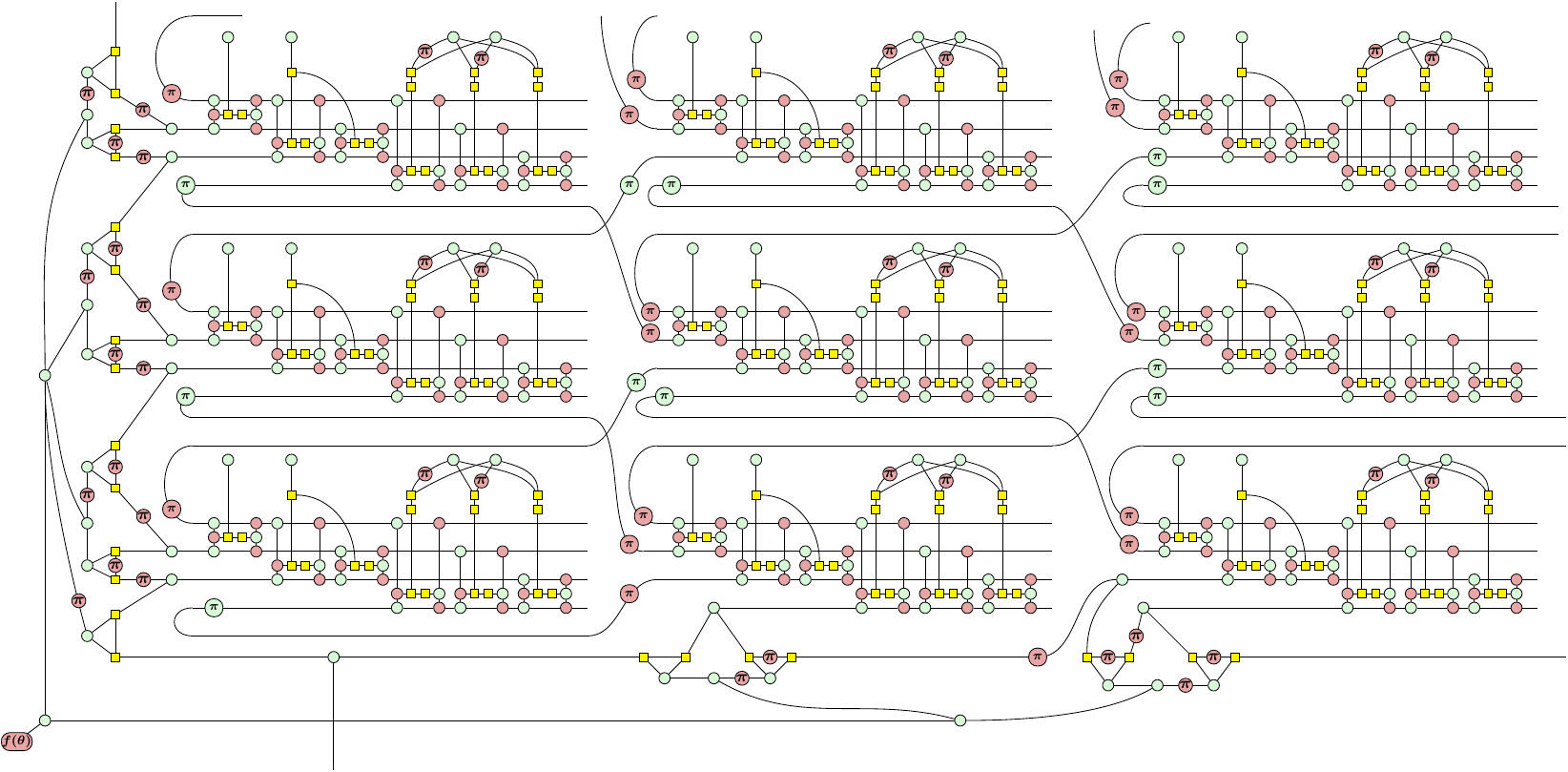}
	\caption{A demonstration of a diagrammatic symmetry transition. When $f(\theta)=0$ (in the bottom-left corner) we recover the mirror-symmetric state of Fig.~\ref{fig:HOTISPT}(a,b), while if $f(\theta)=\pi$ we recover the $C_4$-symmetric state of Fig.~\ref{fig:HOTISPT}(c,d).\label{fig:sym2D} 	}
	\label{fig:AKLT-mps}
\end{figure*}

For $f(\theta)=0$ or $f(\theta)=\pi$ we can start to apply \CopyRule to push the corresponding X spider through the diagram, where it encounters some $\pi$ phases to toggle its behaviour for that particular singlet.

To summarise, we have shown that if a diagram has crystal symmetries, or is built with elements that respect the symmetry, the state the diagram represents has the same symmetries. This allowed us to build a diagram that interpolates, as a function of a control phase, between two different symmetry-protected higher-order topological phases.

}

\section{Conclusion}\label{sec:conclusion}

We introduced the ZXH-calculus as a new tool to represent and operate with quantum states. Specifically, we showed how to represent the 1D and 2D AKLT states as ZXH-diagrams. Using the ZXH-calculus we showed how the non-zero string order of the 1D AKLT state emerges in the ZXH representation, and how to reduce the 2D AKLT state to a graph state using a suitable measurement. 
{ We found two further examples where the diagrammatic nature of the ZXH representation offers an advantage. 
First, we found an explicit expression for the Berry phase of a finite AKLT chain.
Second, we observed that crystal symmetries can be implemented by constructing symmetric diagrams. This observation allowed us to exemplify a transition between a mirror-symmetric and $C_4$ symmetric higher-order topological phase.
In addition, many of the diagrammatic calculations (the entirety of sections \ref{sec:AKLT-in-ZXH} and \ref{sec:2DAKLT}) were} presented solely for pedagogical purposes and are of such a mechanical nature that they can be done straightforwardly by \textsc{PyZX}, a \textsc{Python} package that can simplify ZXH-diagrams.
In the process of constructing the AKLT states, we also found a general way to represent the symmetrising projector on a tensor product of qubit Hilbert spaces in the ZXH-calculus.

Our work opens several directions for further research. One is to seek ZXH-representations of more general quantum states that would allow computations on them to be simplified. 
The success in representing AKLT-type states suggests that more general resonating valence bond states~\cite{Anderson73}, as well as fractional quantum Hall states~\cite{Arovas88} have useful representations in the ZXH-calculus. 
{
Another natural direction is to construct more elaborate symmetry-protected topological phases~\cite{Pollmann2010,Schuch2011,Chen2011,senthil2015}. 
For example, other higher-order topological phases could be built based on the coupled wire construction~\cite{Meng19}, which consists of piling coupled $d-1$-dimensional states together to construct $d$-dimensional topological states.}
More generally, it would be desirable to explore representations of chiral topological states using the ZXH-calculus, especially those with gapped bulk excitations, as those have been challenging to study as PEPS~\cite{Orus:2019vt}. 
{However, it is worth remembering that injective MPS or PEPS cannot represent topological order~\cite{P_rez_Garc_a_2008,P_rez_Garc_a_2010}, a restriction which might present so-far unexplored complications in the ZXH formalism.
}
Interestingly, our discussion of symmetries suggests that a chiral ZXH diagram would necessarily represent a chiral state. However, finding which chiral ZXH diagrams possess non-trivial topological order remains an open question. 
One possible example of a chiral phase is one that would be realised by stacking $C_4$ symmetric 2D-HOTIS in Fig.~\ref{fig:HOTISPT} in such a way that for each consecutive layer the dangling spin-$1/2$ has moved one site along the edge. This state would have $C_4$ symmetry in each plane, but the end states will spiral in the $z$ direction, defining a 3D chiral state with no mirror symmetry. 
Lastly, it is also worth investigating if the  ZXH-calculus allows us to represent and apply matrix product operators (MPO) more efficiently, which are central to MPS-based algorithms~\cite{White1992,White1993,Schollwock11,10.21468/SciPostPhysLectNotes.5}. 
For example, it might be possible for the MPO to be `compressed' using ZXH-calculus rewrite rules.
Lastly, since the ZXH-representation of higher-dimensional Hilbert spaces are not conceptually different, the ZXH-calculus offers a practical way to describe and reason about a broad number of systems, not restricted to one or two spatial dimensions.

More intriguingly, the versatility of the ZXH-calculus could inspire the search for simpler algorithms to tackle many-body problems. For example, it is in principle possible to formulate existing algorithms, such as the density matrix renormalisation group~\cite{White1992,White1993,10.21468/SciPostPhysLectNotes.5}, in terms of ZXH-diagrams and use rewrite rules to reduce the complexity of the involved mathematical objects. 
{ It is interesting to note that our Berry phase proof employed the derivative of a diagram. This suggests that implementing a variational principle based on minimising over a parameter could be feasible. It would overlap with work using these diagrams to analyse problems in quantum machine learning~\cite{zhao2021analyzing,toumi2021diagrammatic}}.
Simplifying to logically equivalent, but representationally simpler objects could be a way to reduce the number of variational parameters. { However, it is an open question how to implement an algorithm that benefits explicitly from the rewrite rules. That said, given successes in quantum compilation algorithms~\cite{kissinger2020reducing,cowtan2020generic,deBeaudrap2020Techniques,hanks2019effective,duncan2019graph} and the aforementioned work on diagrammatic quantum machine learning, we are optimistic.}

In summary, we have presented how the ZXH-calculus can significantly enhance the scope of diagrammatic reasoning to solve many-body quantum problems. 
Our work promotes the helpful pictorial representations of tensor-networks to full-fledged methods of computation, where the diagram \emph{is} the calculation.

\section*{Acknowledgments}
NC was funded by EPSRC fellowship EP/S00114X/1. AGG is indebted to C. Repellin, F. Pollmann, and M. A. S\'{a}nchez-Mart\'{i}nez for enlightening discussions, and acknowledges funding from the French National Research Agency through the project ANR-18-CE30-0001-01 (TOPODRIVE). RDPE would like to acknowledge financial support from the ``Investissements d'avenir'' (ANR-15-IDEX-02) program of the French National Research Agency and discussions with P. Martin-Dussaud. 
JvdW is funded by a NWO Rubicon personal fellowship.

%

\appendix

\ifappendix

\section{Spin matrices and representation theory}\label{app:spin}
The Hilbert space of a spin-chain with $N$ spins is a tensor product of the Hilbert space for each individual spin $s$: $(\mathbb{C}^{2s+1})^{\otimes N}$. For the spin $s=1$ chain this is simply $(\mathbb{C}^{3})^{\otimes N}$.
At each site, the spin-1 matrices that can be used to construct the AKLT Hamiltonian Eq.~\eqref{eq:AKLTham} in the main text can be taken to be
\begin{eqnarray}
   S^{x}&=&\dfrac{1}{\sqrt{2}}
    \left(\begin{array}{ccc}
      0   & 1 & 0 \\
      1   & 0 & 1 \\
      0 & 1 & 0
    \end{array}\right), \hspace{1mm} S^{y}=\dfrac{i}{\sqrt{2}}
    \left(\begin{array}{ccc}
      0   & -1 & 0 \\
      1   & 0 & -1 \\
      0 & 1 & 0
    \end{array}\right), \\
    \label{eq:Smatrices}
    S^{z}&=&\dfrac{1}{\sqrt{2}}
    \left(\begin{array}{ccc}
      1   & 0 & 0 \\
      0   & 0 & 0 \\
      0 & 0 & -1
    \end{array}\right),
\end{eqnarray}
which can be used to define a spin vector at each site $\vec{S}_i=(S^{x}_i,S^{y}_i,S^{z}_i)$. The spin operator $S^a_i$ at site $i$ acts on the local Hilbert space of the $i$-th spin, and thus acts trivially on the full Hilbert space:
\begin{equation}
    S^{a}_{i} = I\otimes I\otimes I \cdots I\otimes I\otimes S^{a}\otimes I \otimes \cdots.
\end{equation}
Hence, for two sites $i,j$ we have the commutation rules
\begin{equation}
    \left[S^{a}_i,S^{b}_j\right]=i\delta_{ij}\epsilon_{abc}S^{c}_j,
\end{equation}
where latin letters label Cartesian directions (e.g., $a=x,y,z$).

Using representation theory we can explain how a spin-1 particle can be decomposed into the symmetric space of two spin-1/2 particles. 
First, recall that we can decompose the four dimensions of the Hilbert space of two spin-1/2 particles into the \emph{triplet representation}, which is spanned by $\ket{00},\frac{1}{\sqrt{2}}(\ket{01}+\ket{10}),\ket{11}$, and the \emph{singlet representation} $\frac{1}{\sqrt{2}}(\ket{01}-\ket{10})$.
Viewing the triplet representation as a three-dimensional Hilbert space, these three spin-1/2 pairs have eigenvalues $s_z = 1,0,-1$ respectively, and so we can view them as a representation of a spin-1.

In general, the tensor product of the Hilbert space of two spins, $s_1$ and $s_2$, can be decomposed into the representations with spins $|s_1-s_2|,|s_1-s_2|+1,\cdots,s_1+s_2$. We can then express the triplet and singlet decomposition for two spin-1/2 particles as 
\begin{equation}
    (1/2)\otimes (1/2) = (0) \oplus (1),
\end{equation}
which is sometimes called a \emph{fusion rule}. 
For two spin-1 particles we get using this rule
\begin{equation}
    \label{eq:deco}
    (1)\otimes (1) = (0) \oplus (1) \oplus (2).
\end{equation}
Note that the only way to get $(2)$ is from $(1)\otimes(1)$. We can use of this property to find the ground state of the AKLT Hamiltonian by expressing the Hamiltonian as a sum of projectors onto the $s=2$ subspace. 

A projector $P^{(s)}$ has eigenvalue $1$ when applied to a state with spin $s$ and zero otherwise. A projector into $m$ spins of total spin $s$ can be built from products of the operator $\hat{O}_j=(\sum_{i}^{m}\vec{S}_{i})\cdot(\sum^{m}_{i}\vec{S}_{i})-j(j+1)$ where $j\neq s$. 
This can be seen by noticing that $S^2\ket{s,s_z}=s(s+1)\ket{s,s_z}$, and thus $\hat{O}_j$ returns zero when applied to a state with total spin $j$.

For two ($m=2$) spin-1 particles, the projector to $s=2$ is constructed by projecting out the $s=0$ and $s=1$ subspaces (choosing $j=0,1$)
\begin{eqnarray} 
 &P^{(2)}(\vec{S}_{1},\vec{S}_{2}) = \lambda\hat{O}_0(\vec{S}_{1},\vec{S}_{2})\hat{O}_1(\vec{S}_{1},\vec{S}_{2}) \\ \nonumber
 &= \lambda [(\vec{S}_{1}+\vec{S}_{2})^2-0(0+1)][(\vec{S}_{1}+\vec{S}_{2})^2-1(1+1)].
\end{eqnarray}
The projector $P^{(2)}$ onto spin-2 annihilates any state with total spin $s$ equal to $0$ or $1$, i.e.~$P^{(2)}\ket{s=1,s_z}=P^{(2)}\ket{s=0,s_z}=0$, where $s_z$ denotes the eigenvalue of the state for $S^z$.
The coefficient $\lambda$ is fixed by the requirement that $P^{(2)}\ket{s=2,s_z}=\ket{s=2,s_z}$ which results in $1/\lambda = [2(2+1)-0][2(2+1)-1(1+1)]=24$. 
By using that $\left(\vec{S}_{1}+ \vec{S}_{2}\right)^2 = \vec{S}^2_{1}+\vec{S}^2_{2}+2\vec{S}_{1}\cdot\vec{S}_{2}$ 
and that $\vec{S}_{1}^2=\vec{S}^2_{2}=2$ for spin-1 we have
\begin{eqnarray}
\nonumber
 P^{(2)}(\vec{S}_{1},\vec{S}_{2}) &=& \dfrac{1}{24} [4+2\vec{S}_{1}\cdot\vec{S}_{2}][2+2\vec{S}_{1}\cdot\vec{S}_{2}]\\
 &=& \dfrac{1}{6}(\vec{S}_{1}\cdot \vec{S}_{2})^2+\dfrac{1}{2}\vec{S}_{1}\cdot \vec{S}_{2}+\dfrac{1}{3}.
\end{eqnarray}
As a result, the AKLT Hamiltonian can be written as
\begin{eqnarray}
\label{eq:AKLThamapp}
    H &=& \sum_{i} \vec{S}_i \cdot\vec{S}_{i+1} + \dfrac{1}{3}  (\vec{S}_i \cdot\vec{S}_{i+1})^2 \\
    &=& 2\sum_{i} \left(P^{(2)}(\vec{S}_{i},\vec{S}_{i+1})- 1/3\right). 
\end{eqnarray}
As we observed below Eq.~\eqref{eq:deco}, the only way for two spin-1 particles to be in the $s=2$ subspace is for each to be in $s=1$ subspace.
Since the AKLT Hamiltonian is the sum of projectors onto the spin-2 subspace of neighbouring spins, it annihilates any state where any two of the four neighbouring spin-1/2 degrees of freedom are in a spin-singlet, because such states have total spin $s=0$. 

Lastly, as mentioned in the main text, the AKLT state has a dilute anti-ferromagnetic order (a site with $s_z=\pm1$ is followed by a site $\mp1$, with a string of $s_z=0$ in between), as discussed in the main text. It can be shown that this order is captured by a non-zero string-order parameter~\cite{Nijs89}.

\section{Additional diagrammatic proofs} \label{Appendix:XandZ-node-extraction}

{
\subsection{Additional proofs for the AKLT Berry Phase calculation}\label{app:Berry-phase-elements}
The following proofs are used in Sec.~\ref{sec:Berry-phase} to derive the Berry phase of the 1D AKLT state. We only use the standard rewrite rules of Figs.~\ref{fig:zx-rules} and~\ref{fig:zh-rules}.
\begin{equation} \label{equation:Chain-top-proof}
	\scalebox{0.8}{%
%
}
\end{equation}
}

\subsection{CSWAP POVM calculations}\label{app:povm-calc}
In the main text it was shown that if the $E_x$ operator is applied to a CSWAP, that the CSWAP is absorbed (see~\eqref{eq:x-povm-simp}).
In this appendix we will show the same for $E_z$ and $E_y$.
First, with $E_z$:
\begin{equation}
	\centering
	\scalebox{0.8}{%
%
%
}
\end{equation}

For the analogous derivation with $E_y$ we need a couple more types of rewrites.
First, there is a way to commute a $\pi$ X-phase through an H-box:
\begin{equation}\label{eq:CZ-rule}
%
%
\end{equation}
This can be proven easily using \SpiderRule to unspider the $\pi$ phase, followed by \HCompRule and \AbsorbRule.

Second, there are ways to remove $\frac\pi2$-labelled Z-spiders and $\pi$-labelled Z-spiders from a diagram, by complementing the connectivity of their neighbours in a suitable way. These were proven in~\cite{duncan2019graph}. To write them down clearly we adopt the notation of \emph{Hadamard-edges} from~\cite{duncan2019graph}:
\begin{equation}
%
%
\end{equation}
The first rule is known as \emph{local complementation}:
\begin{equation}\label{eq:lc-simp}
%
%
\end{equation}
Note that on the right-hand side the middle spider is removed, at the cost of introducing edges between all its neighbours. Because of~\eqref{eq:hopf-ZZ}, if there was already an edge present between the spiders, the edge is cancelled, hence the name complementation.

The second rule is known as \emph{pivoting}:
\begin{equation}\label{eq:pivot-simp}
  \scalebox{0.83}{%
%
%
}
\end{equation}
Here the connected pair of spiders $u$ and $v$ which have a phase of $0$ or $\pi$ are removed on the right-hand side, at the costs of introducing edges between the exclusive neighbourhood of $u$, the exclusive neighbourhood of $v$ and the joint neighbourhood, labelled by respectively $U$, $V$ and $W$ in the diagram.

Now we have all the ingredients we need to prove that $E_y$ applied to the symmetriser reduces to just $E_y$:

\begin{widetext}

\begin{equation}\label{equation:E_y-derivation}
	\scalebox{0.95}{%
%
%
}
\end{equation}

\end{widetext}

While this derivation is significantly more complicated, note that PyZX still manages to simplify it in an automated way (using a different rewrite strategy).

\subsection{Removing \texorpdfstring{$\pi$}{pi} phases from a graph state}\label{app:pi-extraction}

In the main text it was shown how the $\pi$ phases from the singlets on the measured 2D AKLT lattice can be moved onto the external wires for the measurement outcome $E_z$. Here we will demonstrate the same for $E_x$ and $E_z$.

For an $E_x$ outcome in the bulk of the lattice we have: 
\begin{equation}
	\centering
	\scalebox{0.8}{%
%
}
\end{equation}
For an $E_y$ outcome, again in the bulk, we have: 
\begin{equation}
	\centering
	\scalebox{0.8}{%
%
%
}
\end{equation}

{ 
\section{Constructing higher spins in ZXH}\label{app:higher-spins}

In section \ref{sec:higher-spins} we discussed how one can in principle construct spaces for higher spins by making use of CSWAP operators. We then demonstrated the principle for spin 1 and spin-3/2. We here outline how to construct the diagrams for the spin-4 (which is used in section \ref{section:symmeries}) and spin-5/2 symmetrisers. The construction of these symmetrisers show how we can build them for all higher spins.

Recall that we construct the symmetriser on $n+1$ wires $P_S^{(n+1)}$ by making use of the symmetriser $P_S^{(n)}$ and then inserting additional CSWAPs that are fired in a superposition.

For $n=3$ we saw this gives the following diagram:
\begin{equation*}
\centerline
{\scalebox{0.8}{%
%
}}
\end{equation*}
This works, because
\begin{equation*}
%
%
 \ \ = \ \ 2( \ket{00} + \ket{10} + \ket{01}),
\end{equation*}
so that the the latter two CSWAPs generate a superposition of the identity, SWAP$_{1,3}$ and SWAP$_{2,3}$.

For $n=4$ things are slightly more involved. The following diagram is an example of how we can construct it:
\begin{equation}
	\centerline
	{\scalebox{0.9}{%
%
%
}}
	\label{diagram:4-symmetriser}
\end{equation}

For $n=5$ we can use the following diagram:
\begin{widetext}
\begin{equation}\label{diagram:5-symmetriser}
	{\scalebox{1.0}{%
%
%
}}
\end{equation}
\end{widetext}

The goal of the `crowns' over these operators is to generate a superposition over all states that fire at most one of the CSWAPs. Recall that the swap gates are triggered when fed a $\ket{1}$, corresponding to an X $\pi$ phase. So the desired superposition consists of those states $\ket{x_1\cdots x_n}$ where at most one of the $x_i$ is a $1$. For example, for $n=4$ we want $\ket{000}+\ket{001}+\ket{010}+\ket{100}$. The first step in creating this state in ZXH is to take $k$ Z-spiders such that $2^k\geq n$. For $n=4$ we need two (arity 3) green spiders. These spiders give us:
$$
\ket{000}\ket{000}+\ket{000}\ket{111}+\ket{111}\ket{000}+\ket{111}\ket{111}.
$$
We then use NOT gates (X $\pi$ phases) and AND gates (pairs of H-boxes and Hadamards) to transform each of these four terms into one of the states we want. In~\eqref{diagram:4-symmetriser} for instance, the three AND gates `select' the state $\ket{01}+\ket{10}+\ket{11}$. The superposition also contains a $\ket{00}$, but this is not passed through the ANDs, and hence results in none of the CSWAPs firing, which is the final state we require.

The $n=4$ case is special, as in general there will be redundant states in the superposition. If we consider the crown of~\eqref{diagram:5-symmetriser} we see that we have $2^3$ states in superposition but we only require five of them:
$$
\ket{000}+\ket{001}+\ket{010}+\ket{100}+\ket{111}.
$$
As a result we must `bin' the rest of the states. This is done by using an AND gate that is post-selected to a $\ket{0}$ outcome. We can represent such a post-selected AND by an H-box with a Z-spider attached to it (see the top of the diagram). Indeed, looking at the crown of~\eqref{diagram:5-symmetriser} we see that it selects four elements of the superposition to trigger gates (which precise ones it selects is not irrelevant). It then discards three more states leaving one state ($\ket{111}$) left over that is present in the superposition, but triggers no additional CSWAPs so that we get the required identity gate. This construction for $n=5$ generalises to any desired $n$.

}

 \clearpage
 \begin{widetext}

 \section{Overview of graphical rewrite rules}\label{app:overview-rules}
 \begin{align*}
 &{} \scalebox{1.0}{%
%
 \end{equation*}

 \end{widetext}

\fi

\end{document}

%% file: ZX.tikzstyles
\tikzstyle{Green Node}=[fill={zx_green}, inner sep=0mm, minimum size=2mm, draw=black, shape=circle, tikzit fill=green, font={\footnotesize\boldmath}]
\tikzstyle{Red Node}=[fill={zx_red}, inner sep=0mm, minimum size=2mm, draw=black, shape=circle, tikzit fill=red, font={\footnotesize\boldmath}]
\tikzstyle{H}=[fill=yellow, draw=black, shape=rectangle, inner sep=0.6mm, minimum height=1.5mm, minimum width=1.5mm]
\tikzstyle{new style 0}=[fill={rgb,255: red,255; green,0; blue,4}, draw=black, shape=rectangle]
\tikzstyle{new style 1}=[fill=green, draw=black, shape=rectangle]
\tikzstyle{swirl}=[fill=white, draw=black, shape=circle]
\tikzstyle{new style 2}=[fill=black, draw=black, shape=circle]
\tikzstyle{gphase}=[rounded rectangle, rounded rectangle arc length=90, fill={zx_green}, inner sep=2pt]
\tikzstyle{rphase}=[rounded rectangle, rounded rectangle arc length=90, fill={zx_red}, inner sep=2pt]
\tikzstyle{box}=[shape=rectangle, text height=1.5ex, text depth=0.25ex, yshift=0.5mm, fill=white, draw=black, minimum height=5mm, yshift=-0.5mm, minimum width=5mm, font={\small}]
\tikzstyle{Z dot}=[Green Node, tikzit fill=green]
\tikzstyle{Z phase dot}=[minimum size=4mm, font={\footnotesize\boldmath}, shape=rectangle, rounded corners=1.5mm, inner sep=0.2mm, outer sep=-2mm, scale=0.8, tikzit shape=circle, draw=black, fill={zx_green}, tikzit draw=blue, tikzit fill=green]
\tikzstyle{X phase dot}=[Z phase dot, tikzit shape=circle, fill={zx_red}, font={\footnotesize\boldmath}, tikzit draw=blue, tikzit fill=red]
\tikzstyle{X dot}=[Red Node, tikzit fill=red]
\tikzstyle{hadamard}=[H, tikzit shape=rectangle, tikzit fill=yellow]
\tikzstyle{vertex}=[inner sep=0mm, minimum size=1mm, shape=circle, draw=black, fill=black]
\tikzstyle{vertex set}=[inner sep=0mm, minimum size=1mm, shape=circle, draw=black, fill=white, font={\footnotesize\boldmath}]

\tikzstyle{Edge}=[-]
\tikzstyle{new edge style 0}=[draw=black, {|->}]
\tikzstyle{new edge style 1}=[-, draw={rgb,255: red,191; green,0; blue,64}]
\tikzstyle{brace edge}=[-, tikzit draw=blue, decorate, decoration={brace,amplitude=1mm,raise=-1mm}]
\tikzstyle{hadamard edge}=[-, color=blue, dashed, dash pattern=on 3pt off 1.5pt, thick, tikzit draw=blue]
\tikzstyle{dashed gray}=[-, dashed, tikzit fill={rgb,255: red,191; green,191; blue,191}, fill={rgb,255: red,191; green,191; blue,191}, draw={rgb,255: red,128; green,128; blue,128}]